\documentstyle[graphics,psfig]{aa}

\begin{document}

\title{XMM-Newton EPIC observations of 21 low-redshift PG quasars
}
\author{D. Porquet\inst{1} \and J.N. Reeves\inst{2,3} 
\and P. O'Brien\inst{4}\and W. Brinkmann\inst{1} 
}

\offprints{D. Porquet}
\mail{dporquet@mpe.mpg.de}

\institute{
Max-Plank-Institut f\"{u}r extraterrestrische Physik, Postfach 1312,
D-85741, Garching, Germany
\and Laboratory for High Energy Astrophysics, 
NASA Goddard Space Flight Center, Greenbelt, MD 20771, USA
\and Universities Space Research Association
\and  Department of Physics and Astronomy, 
University of Leicester, Leicester LE1 7RH, UK 
}

\date{Received January 21, 2004 / Accepted April 8, 2004}

\abstract{We present an X-ray spectral analysis of 21 low redshift 
  quasars observed with {\sl XMM-Newton} EPIC. 
All the sources are Palomar Green quasars with redshifts between 0.05 and 
0.4 and have low Galactic absorption along
 the line-of-sight.  A large majority of quasars in the sample (19/21) 
exhibit a significant soft excess below $\sim$  1--1.5\,keV, whilst 
 two objects (PG\,1114+445 and I\,Zw1) show a deficit of soft X-ray flux 
due to the presence of a strong warm absorber. 
Indeed, contrary to previous studies with {\sl ASCA} and {\sl ROSAT}, 
we find that the presence of absorption features 
near  0.6--1.0\,keV is common in our sample.
  At least half of the objects appear to harbor a warm absorber, 
as found previously in Seyfert 1 galaxies.
 We find significant detections of Fe\,K$\alpha$ emission lines in at least 
twelve objects, whilst there is evidence for some broadening of the line 
profile, compared to the EPIC-pn resolution, in five of these quasars. 
 The determination of the nature of this broadening (e.g., 
Keplerian motion, a blend of lines, relativistic effects) is not possible 
with the present data and requires either higher S/N or higher resolution 
spectra. 
In seven objects the line is located between 6.7--7\,keV, corresponding 
to highly ionized iron, whereas in the other five objects the line energy 
is consistent with 6.4 keV, i.e. corresponding to near neutral iron. The 
ionized lines tend to be found in the quasars with the steepest 
X-ray spectra.  
We also find a correlation between the continuum power law index $\Gamma$ and 
the optical H$\beta$ width, in both the soft and hard X-ray bands, 
whereby the steepest X-ray spectra are found in objects with narrow 
H$\beta$ widths, which confirms previous {\sl ROSAT} and  {\sl ASCA} results. 
The soft and hard band X-ray photon indices are also strongly correlated, 
i.e. the steepest soft X-ray spectra correspond the steepest
 hard X-ray spectra.
We propose that a high accretion rate and a smaller black hole mass 
is likely to be the physical driver responsible for these trends, 
with the steep spectrum objects likely to have smaller black hole 
masses accreting near the Eddington rate.

}
\titlerunning{XMM-Newton observations of 21 low-z PG quasars}
\authorrunning{Porquet et al.}
\maketitle
\section{Introduction}

In Active Galactic Nuclei (AGN), from Seyfert galaxies to quasars, 
 the analysis of spectral X-ray features can help us to understand
 the central regions of these powerful objects. 
The first is the so-called soft excess
seen below 2--3\,keV (Arnaud et al. \cite{Arnaud85}; 
Turner \& Pounds \cite{Turner89}).
This spectral characteristic is thought to be the high energy part of the 
  optical-UV ``big blue bump''  extending down to 1$\mu$m, 
which contains a large fraction 
of the bolometric luminosity. 
Soft X-ray excesses were detected for most AGN with {\sl ROSAT} (e.g., Brinkmann \cite{Br92}, Buehler et al. \cite{Bu95}).  
 Using {\sl XMM-Newton} observations, Pounds \& Reeves (\cite{Pounds2002}),
 recently confirmed  the presence of a soft X-ray excess in all objects of their sample (6 Seyfert galaxies), 
with  amplitude and width increasing with luminosity. 
Current interpretations of the soft excess range from direct thermal 
emission from the accretion disk to reprocessing of harder radiation absorbed 
in the thin disk (Pounds \& Reeves \cite{Pounds2002}).
 Other important components are emission and/or absorption structures,
mainly in the soft X-ray range, due to a warm absorbing  
medium supposed to be located between the  Broad Line Region 
and the Narrow Line Region 
(e.g., Reynolds \& Fabian \cite{RF95}; Porquet et al. \cite{Porquet99}). 
 These absorption and/or emission features are commonly  
seen in low-luminosity AGN such as Seyfert galaxies, 
 but are rather rare in quasars. 
The Fe\,K$_{\alpha}$ line complex observed in the 6--7\,keV range is
also an important spectral diagnostic tool to probe dense matter
 from the inner disk  (e.g., \object{MCG-06-30-15}, Tanaka et al. \cite{T95},
 Fabian et al. \cite{F2002}; \object{Q0056-363}, Porquet \& Reeves \cite{PR2003})
to the Broad Line Region and the molecular torus 
(e.g.  \object{NGC 5548}, Yaqoob \cite{Ya2001};
 \object{Mrk 205}, Reeves et al. \cite{R2001};
\object{Mrk 509}, Pounds et al. \cite{Pounds2001}). 
  Recently, Page et al. (\cite{Pa2004}) have shown that 
the majority of 53 AGN observed with {\sl XMM-Newton}, 
including both Seyfert 1 and quasars, exhibit narrow Fe\,K line near 6.4\,keV, 
due to the emission from neutral material, such as the BLR or the 
putative molecular torus. \\
\indent In this paper,  we present the spectral analysis 
of a sample of 21 low-redshift  quasars (z$\sim$0.05--0.4),
obtained with {\it XMM-Newton}. 
 One great advantage of such study with the EPIC cameras (MOS and pn) 
is the combination of high sensitivity with moderate spectral resolution over 
a broad energy band, i.e. from 0.3\,keV up to 12\,keV for the pn camera. 
All sources are Palomar Green objects (Green et al. \cite{Green86}) 
with low absorption column density  along the line-of-sight.
Note that all fit parameters given in this paper are in the quasars rest frame, 
with values of H$_{\rm 0}$=75\,km\,s$^{-1}$\,Mpc$^{-1}$, 
and q$_{\rm 0}$=0.5 assumed throughout. \\
\indent In section~\ref{sec:xmm}, we present the quasar sample used in this work 
as well as the data reduction techniques. 
 In section~\ref{sec:mdot}, we estimate the accretion rate 
with respect to the Eddington rate for each quasar. 
In section~\ref{sec:xmmspectra}, we analyze the spectral shapes
 over different energy bands. 
The possible presence of a warm  absorber medium and of 
a Fe\,K line is checked in sections~\ref{sec:WA} and \ref{sec:FeK},
 respectively. 
 Then, in sect.~\ref{sec:all}, we investigate possible potential origins  
for the soft excess and which model could account for the whole 
 spectral shape. Finally, in sect.~\ref{sec:corr}, 
we report the global properties of the quasar sample
 and check for correlations between the X-ray properties
 and other quasar properties. 
In appendix~A, we report individual notes
 about each object of our sample.

\section{XMM-Newton observations}\label{sec:xmm}

\subsection{The sample}

We present archival {\sl XMM-Newton} observations 
 of 21 low-redshift quasars. 
In Table~\ref{tab:sample}, we report for each object:
 coordinates; 
 redshift from Schmidt \& Green (\cite{SG83}); 
 Galactic column density along the line-of-sight (N$_{\rm H}^{\rm Gal}$);
 visual absolute  magnitude  (M${\rm v}$) from Boroson \& Green (\cite{BG92}); 
  ratio of radio to optical luminosity  
(radio-loudness R$_{\rm L}$, in logarithmic scale) as defined
 in  Kellerman et al. (\cite{Kel89}); 
 Full Width at Half Maximum (FWHM) of the H$_{\beta}$ line 
at 5100\AA~ from Boroson \& Green (\cite{BG92}), 
and the object type.
The Galactic column density values have been obtained from the {\sc coldens} 
program using the compilations of Dickey \& Lockman (\cite{DL90}). 
Five objects can be classified as narrow line galaxies (NLG)
 with FWHM~$\leq $~2\,000 \,km\,s$^{-1}$: 
 I\,Zw1,  PG\,1115+407, PG\,1244+026,  PG\,1402+261, and Mrk 478;  
and two are radio-loud quasars (RLQ) with R $\geq$ 1.0: PG\,1309+355,
 and PG\,1512+370.
The other 14 objects are radio-quiet quasars (RQQ). 
 The Palomar Bright Quasar Survey (BQS) contains 114 objects
 (see Kellerman et al. \cite{Kel89}), 
including 72 quasars with redshift between 0.048 and 0.400. 
Taking the values of M${\rm v}$ and FWHM(H$_{\beta}$) 
from Boroson \& Green (\cite{BG92}), 
we found for these 72 objects that the values span a large range:  
 $-$21.43 $\leq$ M${\rm v}$ $\leq$ $-$27.15 
and 830 $\leq$ FWHM(H$_{\beta})$ $\leq$ 9410 km\,s$^{-1}$.   
The corresponding average values are $-$24.3 ($\sigma$=1.2) 
and 3712 \,km\,s$^{-1}$ ($\sigma$=1980 km\,s$^{-1}$) respectively.  
The present 21 objects represent about one third of the latter objects, 
and the values of M${\rm v}$ and FWHM(H$_{\beta}$) also  
 cover a large range: $-$21.77 $\leq$ M${\rm v}$ $\leq$ $-$25.93 
and 830 $\leq$ FWHM(H$_{\beta})$ $\leq$ 8450 km\,s$^{-1}$.   
The corresponding average values are 
 $-$24.2 ($\sigma$=0.9) and 3577 \,km\,s$^{-1}$ ($\sigma$=1933 km\,s$^{-1}$),
 respectively. 
Therefore the present sub-sample can  properly represent the properties 
of the larger sample of low-redshift PG quasars, 
without any obvious bias towards M${\rm v}$ and FWHM(H$_{\beta}$). 

\subsection{Data reduction and analysis}

Table~\ref{tab:obs} reports for each object in the sample: the observation date, 
the corresponding {\sl XMM-Newton} orbit revolution, the observation ID, and the exposure time. 
The EPIC MOS cameras (Turner et al. \cite{Turner2001}) were  
 operated in large window mode,  
 except for I\,Zw1  and  PG\,1307+085 (small window mode),
 and for PG\,1244+026 and Mrk 1383  (full window mode). 
 The EPIC pn camera (Str\"uder et al. \cite{Strueder2001}) was also operated  
  in  Large Window mode, except for PG\,0804+761, PG\,1244+026, Mrk 1383, 
 and Mrk 876  (full window mode).
 For all observations the thin filter was used, except for I\,Zw1 (medium filter). 
The data were  re-processed and cleaned using the 
{\it XMM-Newton} {\sc SAS version 5.4.1} (Science Analysis Software) package. 
Since pile-up effects are negligible,  
X-ray events corresponding to patterns 0--12 and 0--4  (singles and doubles) 
 are selected from the MOS and pn cameras, respectively;
 except in the case of Mrk 1383 
for which we use only single events. 
As reported recently by Kirsch (\cite{K2003}), significant differences occurred 
in the low energy redistribution characteristics of the MOS cameras 
compared to early observations. This means that from approximately revolution 450 onwards 
 the calibration below 500\,eV is suspect for the MOS.
 Therefore for observations with a revolution number above 440, we use 
 a low energy cutoff at 500\,eV for the MOS data, 
and a cutoff at 300\,eV for the other MOS data and for all pn data. 
  Since the pn CCD has higher sensitivity and a broader 
energy range (0.3--12\,keV), 
 we use the time-averaged pn spectrum  for each object for the spectral analysis 
(except for PG\,0804+761 for which only MOS data are usable). 
 Indeed pn spectra are better suited for the detection of Fe\,K line 
and the analysis of the soft excess down to 0.3\,keV. 
 However,  we have checked that the results obtained from the pn data
 are consistent with those from the MOS data. 
 The spectra are binned to give a minimum of 100 counts per bin
 for I\,Zw1, and of 40 counts per bin for 
PG\,0953+414, PG\,1048+342, PG\,1114+445,  PG\,1202+281. 
The other spectra are binned to give a minimum of 20 counts per bin.\\
  The errors quoted correspond to 90$\%$ 
confidence ranges for one interesting parameter 
($\Delta \chi^{2}$=2.71).
 Abundances are those of Anders \& Grevesse (\cite{Anders89}). 
In the following, we use the updated cross-sections for X-ray absorption by 
the interstellar medium  from Wilms et al. (\cite{Wilms2000}).  \\
\indent F-Tests are performed (Bevington \& Robinson \cite{BR92}) 
to estimate the significance of  spectral features 
such as emission lines and absorption edges. 
Only features detected with a significance greater than 90$\%$ are considered.

\section{Quasar accretion rate with respect to the Eddington rate}\label{sec:mdot}

In this section, the accretion rates are  estimated with respect to the Eddington rate, 
 defined  as:
\begin{eqnarray}\label{eq:mdot}
\dot{m}= \frac{{\dot{\rm M}}}{{\dot{\rm M}_{\rm Edd}}}  = 
\frac{\rm L_{\rm bol}}{\rm L_{\rm Edd}} = \frac{\rm L_{\rm bol}}{1.26 \times 10^{38}~ \rm M_{\rm BH}}, 
\end{eqnarray}
where ${\dot{\rm M}}$ and ${\dot{\rm M}_{\rm Edd}}$ are the quasar mass accretion rate 
and the Eddington accretion rate (in M$_{\odot}$\,yr$^{-1}$), respectively; 
L$_{\rm bol}$ and L$_{\rm Edd}$ are the bolometric and Eddington luminosities (in erg\,s$^{-1}$), respectively; 
and M$_{\rm BH}$ is the black hole mass (in M$_{\odot}$). 
In Table~\ref{tab:mdot}, the following values are reported:  
 column (1) the black hole mass (expressed in M$_{\odot}$), 
  (2) the monochromatic luminosity  $\nu$L$_{\nu}$(5100\AA) (in erg\,s$^{-1}$), 
  (3) the bolometric luminosity (in erg\,s$^{-1}$), 
and (4) the   accretion rate with respect to Eddington rate (${\dot{\rm m}}$). 
The bolometric luminosities (in erg\,s$^{-1}$)  are from Woo \& Urry (\cite{WU2002}),
  Grupe et al. (\cite{G2004}), or are estimated by using  the approximation: 
 L$_{\rm bol}\simeq$ 9 $\times$ $\nu$L$_{\nu}$(5100\AA) reported in Kaspi et al. (\cite{K2000}).
Woo \& Urry (\cite{WU2002}) calculated the bolometric luminosities 
  either from spectral energy distribution (SED) fitting or by flux integration. 
Grupe et al. (\cite{G2004}) estimated the bolometric luminosities 
by a combined power law with exponential cut-off to the optical-UV data 
and an absorbed power law to the soft X-ray data.  
The accretion rates reported here are only approximations since 
 both the estimates of the bolometric luminosities and of the black hole
masses are subject to uncertainties. 
The determination of the bolometric luminosity is affected by spectral variability 
 and lack of wavelength coverage, mainly in the UV band near 912\,\AA,  
 where most of the luminosity is emitted, 
 due to absorption by Galactic neutral hydrogen. 
 As shown by Ferrarese et al. (\cite{Fe2001}), 
the black hole masses inferred, from reverberation mapping (mean, rms) 
and  from M$_{\rm BH}$ versus stellar velocity relation, 
are consistent with each other.


\section{XMM-Newton spectra}\label{sec:xmmspectra}

\begin{figure*}[t!]
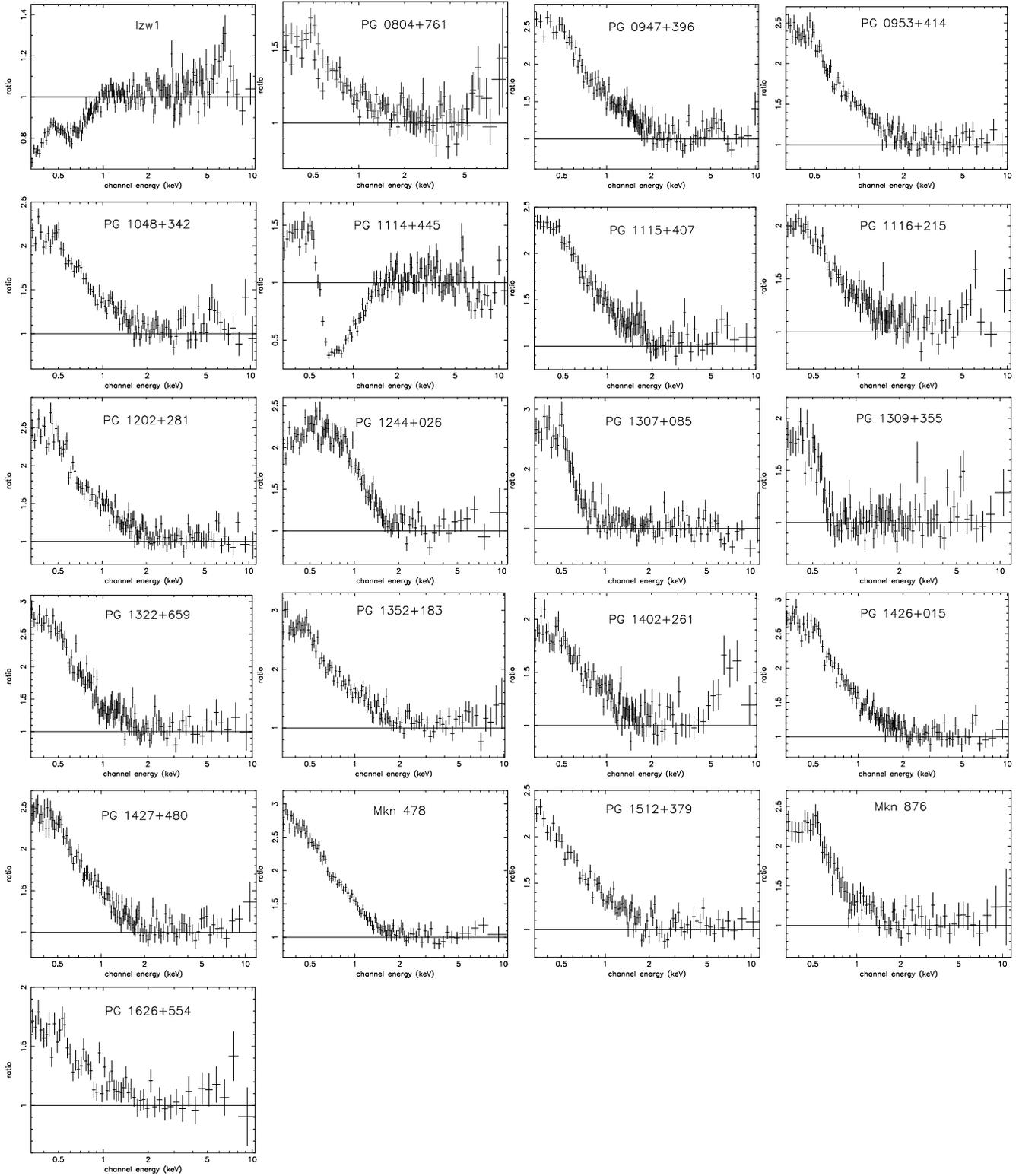

\begin{tabular}{@{}c@{}c@{}c@{}c}
\psfig{file=0108fig1a.ps,width=4.4cm,angle=-90} & \psfig{file=0108fig1b.ps,width=4.4cm,angle=-90} &\psfig{file=0108fig1c.ps,width=4.4cm,angle=-90} &\psfig{file=0108fig1d.ps,width=4.4cm,angle=-90}\\
\psfig{file=0108fig1e.ps,width=4.4cm,angle=-90}  & \psfig{file=0108fig1f.ps,width=4.4cm,angle=-90} & \psfig{file=0108fig1g.ps,width=4.4cm,angle=-90} & \psfig{file=0108fig1h.ps,width=4.4cm,angle=-90} \\
\psfig{file=0108fig1i.ps,width=4.4cm,angle=-90} &\psfig{file=0108fig1j.ps,width=4.4cm,angle=-90} & \psfig{file=0108fig1k.ps,width=4.4cm,angle=-90} &\psfig{file=0108fig1l.ps,width=4.4cm,angle=-90} \\
 \psfig{file=0108fig1m.ps,width=4.4cm,angle=-90} &  \psfig{file=0108fig1n.ps,width=4.4cm,angle=-90} & \psfig{file=0108fig1o.ps,width=4.4cm,angle=-90} & \psfig{file=0108fig1p.ps,width=4.4cm,angle=-90}\\
 \psfig{file=0108fig1q.ps,width=4.4cm,angle=-90} &  \psfig{file=0108fig1r.ps,width=4.4cm,angle=-90} & \psfig{file=0108fig1s.ps,width=4.4cm,angle=-90} & \psfig{file=0108fig1t.ps,width=4.4cm,angle=-90} \\ \psfig{file=0108fig1u.ps,width=4.4cm,angle=-90}\\
\end{tabular}
\caption{Data/model ratio of an absorbed power law fit to the 2--5 keV energy range extrapolated over the 0.3--12\,keV energy range for the 21 objects of the sample  (observer frame).}
\label{fig:spectrum}
\end{figure*}

\indent  In all subsequent fits (except when considering an
absorbed power law fit to the 2--5 keV range where we only use Galactic
absorption),  the column density  is fitted with two components:
 a Galactic column fixed to the values reported in Table~\ref{tab:sample} (N$_{\rm H}^{\rm Gal}$), 
as well as possible intrinsic absorption (N$^{\rm in}_{\rm H}$) in the QSO rest frame, 
which is allowed to vary. \\
\indent A single absorbed power law model over the total 
 0.3--12\,keV energy range gives a poor fit for all  sources in the sample. 
To characterize the hard X-ray continuum, we fit an absorbed 
power law model over the 2--5\,keV energy range where the spectra should
be relatively unaffected by the presence 
of a broad soft excess, of a warm absorber-emitter 
medium, of a Fe\,K$_{\alpha}$ emission line, 
and of a contribution  from a high energy Compton reflection hump above 8\,keV.
In this energy range, all spectra are well fitted by a
 single power law model (Table~\ref{table:fits}).  
 The X-ray emission of the sample in this energy band covers a wide range 
of photon indices from hard
 (e.g.,  $\Gamma\sim$1.4 for PG\,1114+445 and PG\,1307+085)
 to soft (e.g., $\Gamma\sim$2.5 for PG\,1244+026). 
Figure~\ref{fig:spectrum} displays the data/model ratio fit in the 2--5\,keV energy range, 
extrapolated over the 0.3--12\,keV energy band. 
All objects exhibit a positive excess below 1--1.5\,keV, the so-called 
soft excess, except  I\,Zw1, and PG\,1114+445, 
which display very significant negative deviations in the soft part of the spectrum.  
  PG\,1114+445 exhibits a huge absorption feature between 0.6 and 1.5\,keV. \\
\indent Fitting the soft part (0.3--2\,keV) with an absorbed power law model, we
obtain a good fit for 12 objects (see Table~\ref{table:fits}).
The inferred photon indices (2.0 $\leq \Gamma \leq$ 3.1) correspond to steep spectra.  
We show below that for some objects the addition of one or two  absorption edges 
improves the fit in this energy band significantly. \\
\indent Next we fit the data with a broken power law model to take into account 
the soft and hard spectral shapes. 
 As reported in Table~\ref{table:fitbkn}, all objects are well described by
 this continuum model, except for the two Narrow Line type objects 
 I\,Zw1 and  PG\,1244+026, and for the RQQ PG 1114+445. 
For I\,Zw1 and PG 1114+445 this due to the presence of a strong absorption
 feature(s) below 1\,keV. For PG\,1244+026, the soft X-ray shape
 appears more curved than for the other objects (Figure~\ref{fig:spectrum}).

\section{The warm absorber}\label{sec:WA}

The first detection of the signature of a warm absorber (WA) located inside an AGN 
 was the  absorption edges of \ion{O}{vii} (0.74\,keV) and 
\ion{O}{viii} (0.97\,keV) in the quasar \object{MR 2251-158}
 (Halpern \cite{H84}) with the {\sl Einstein} observatory.
Such features were commonly observed in Seyfert\,1 galaxies  
 with  {\sl ASCA} (e.g., Reynolds \& Fabian \cite{RF95}, 
Reynolds \cite{Re97}, George et al. \cite{G98}).
However, this highly ionized medium is not only an absorber but also an emitter
 as predicted by Netzer (\cite{Ne96}), and confirmed by the new generation of 
X-ray satellites {\sl Chandra} and {\sl XMM-Newton}
 which have revealed the very complex soft X-ray spectra of Seyfert galaxies 
(e.g., Kaspi et al. \cite{K2001}, Kaastra et al. 2002, 
Kinkhabwala et al. \cite{Ki2002}). For Seyfert\,1 galaxies, numerous  
absorption features (and a few emission features)
 have been found, while for Seyfert 2 galaxies only emission features
 were detected. This is in  good agreement with the Unified Scheme,
 where in the case of Seyfert 2 galaxies  
the central X-ray source is hidden by a molecular torus, 
thus the WA is only  detected in emission. 
However in quasars, the detection of a WA appeared to be very rare 
(e.g., Reeves \& Turner \cite{RT2000}, George et al. \cite{G2000}), 
and only a few detections have been reported with {\sl ASCA} and {\sl ROSAT}; 
 for example at low redshifts MR 2251-158 (z=0.06) by  Halpern (\cite{H84});   
 and at higher redshifts PG 1114+445 (z=0.144) 
by Reeves \& Turner (\cite{RT2000}), 
 and \object{PDS 456} (z=0.184) by Reeves et al. (\cite{R2003}). 
 Recently with {\sl XMM-Newton}, Brinkmann et al. (\cite{Br2004a}) reported 
absorption and emission features in the soft X-ray part 
of two X-ray weak quasars \object{PG 1411+442} and \object{Mrk 304},
strongly suggesting the presence of ionized material in the central region 
of these two sources.

In our sample,  I\,Zw1 clearly shows absorption features below 1\,keV, 
while PG\,1114+445 has a huge absorption  trough 
between 0.6 and 1.5\,keV as displayed 
in Figure~\ref{fig:spectrum}.  
We now describe the spectral analysis of these features for these two objects, 
 and we also check for such signatures of the WA in the other quasars.
 We first fit the spectra combining an absorbed broken power law with
two absorption edges of \ion{O}{vii} and \ion{O}{viii}. 
For PG\,1114+445, a single underlying power law model is used.
The energies of the edges are initially fixed to their rest 
frame values at 0.74 keV and 0.87 keV respectively
(see results in Table~\ref{tab:WA}). 
 There are six  significant detections above about 99$\%$ confidence. 
Then, in order to allow for a possible outflow or inflow of the WA,
 we leave the energies of the absorption edges as free parameters. 
 We find significant detections ($>$99$\%$ compared to the continuum alone) 
of absorption edges in ten objects: I\,Zw1, PG\,0804+761, PG\,0947+396, 
PG\,1114+445, PG\,1115+407, PG\,1202+281, PG\,1307+085, 
PG\,1309+355, PG\,1352+183,  and Mrk\,876. 
 We do not claim that the absorption features found here
are from oxygen edges but due to the low resolution of EPIC data below 1 keV, 
we fit the data with one or two absorption edges as a simple parameterisation 
only. Indeed 
the detected absorption features could also correspond to absorption lines 
from N, O or Ne as well as Unresolved Transition Arrays 
(UTA, Behar et al. \cite{Be2001}) of absorption from low ionization 
(Fe $<$ XVII) iron L-shell lines 
as recently found in \object{IRAS 13349+2438} with {\sl XMM-Newton}/RGS 
(Sako et al. \cite{Sa2001}). 
 In addition, the presence of narrow absorption lines in some objects 
cannot be inferred with the lower resolution EPIC data presented here.
Therefore, the determination of the exact nature 
(ionization parameter, outflow velocity, etc..) of the WA  
 and of its complexity, as well as the exact fraction of objects
exhibiting a warm absorber, require a much more detailed analysis 
 with the higher spectral resolution of the RGS, which will be presented
 in a forthcoming paper. 
 In conclusion, here we found that at least half of the 21  
objects probably show evidence for absorption features.

\section{The Fe\,K$_{\alpha}$ line}\label{sec:FeK}

\begin{figure*}[t!]
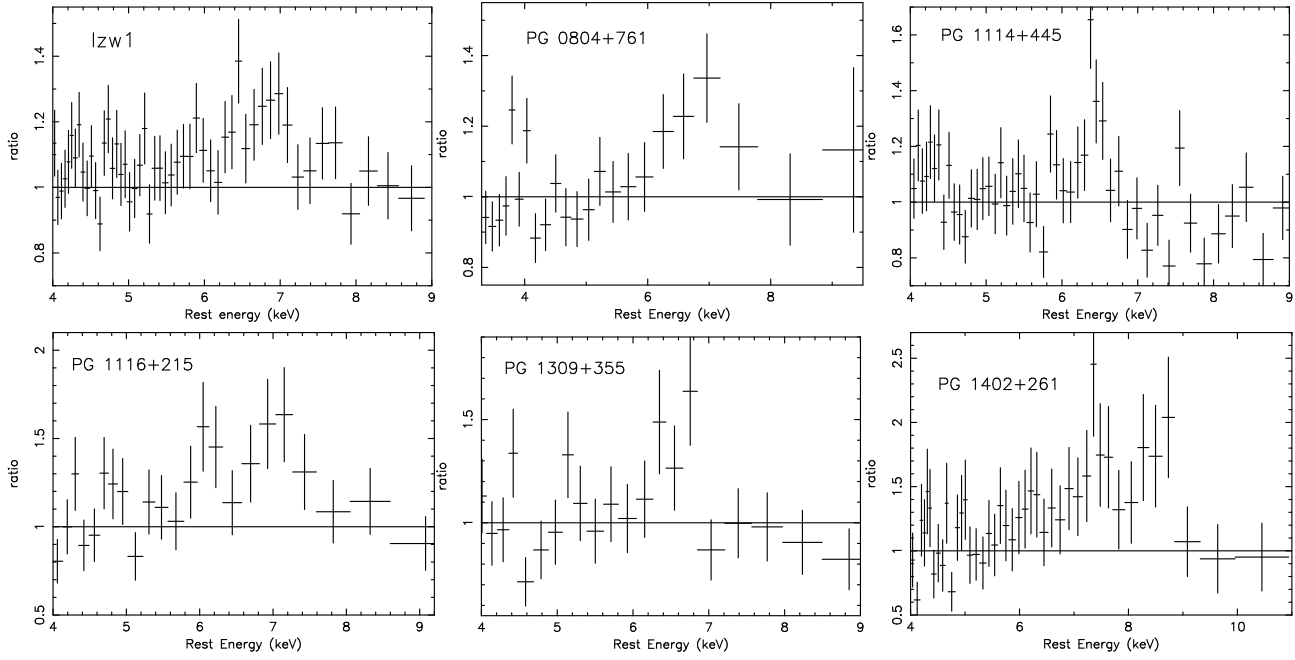

\begin{tabular}{@{}c@{}c@{}c@{}c}
\psfig{file=0108fig2a.ps,width=5.7cm,angle=-90} & \psfig{file=0108fig2b.ps,width=5.7cm,angle=-90} &\psfig{file=0108fig2c.ps,width=5.7cm,angle=-90} \\
\psfig{file=0108fig2d.ps,width=5.7cm,angle=-90}  & \psfig{file=0108fig2e.ps,width=5.7cm,angle=-90} & \psfig{file=0108fig2f.ps,width=5.7cm,angle=-90} \\
\end{tabular}
\caption{Details of the iron line profiles (data/model ratio as in Fig.~\ref{fig:spectrum}, in observer frame) for the six objects which show lines at very high confidence level.}
\label{fig:fezoom}
\end{figure*}

Next we check for the presence of a Fe\,K$_{\alpha}$ line in each 
spectrum. 
 Indeed as shown in Figure~\ref{fig:spectrum} some spectra 
 exhibit a positive deviation at about 6.4--7\,keV compared to the continuum. 
Figure~\ref{fig:fezoom} 
displays a zoom of these features (data/continuum ratio) 
for six objects (I\,Zw1, PG\,0804+761, PG\,1114+445, PG\,1116+215, 
PG\,1309+355, and PG1402+261).
We add a Gaussian line to the absorbed 
 broken power law model in the 0.3--12\,keV energy range 
 (including absorption edges when required), 
except in the case of PG\,1244+026 where we fit a simple power law continuum 
in the 2--12\,keV energy range. 
 Table~\ref{tab:FeK} reports the parameters of the fitted lines. 
F$^{\rm line}_{\rm prob}$ is the ``significance'' of the line compared
 to the continuum model alone  
 (only detections with F$^{\rm line}_{\rm prob}\geq$  90$\%$ are quoted). 
First we fix the width of the line at 10\,eV, 
i.e.  the line is intrinsically narrow, much narrower than the resolution 
of the EPIC-pn at this energy. We find evidence of a Fe K shell emission 
 line in at least twelve objects. 
In five objects the line energy is consistent within the errors  
 to ``neutral'' or moderately ionized iron near 6.4 keV (i.e. $\leq$ \ion{Fe}{XVII}):
 PG\,0947+396, PG\,1048+342, PG\,1114+445, PG\,1309+355, and PG\,1512+379.  
 For the other seven objects the line energy is consistent with highly ionized iron: 
 I\,Zw1, PG\,0804+761, PG\,1115+407, PG\,1116+215, PG\,1244+026, PG\,1402+261, 
and Mrk\,1383. 
We notice that for PG\,0947+396,  PG\,1115+407, PG\,1244+026, 
PG\,1309+355, Mrk\,1383, and PG\,1512+379 
the detections are less significant, between 90--99\% confidence.  \\
\indent  For  five objects the fit is statistically improved 
(F$^{\rm broad}_{\rm prob}\geq$  90$\%$) 
  allowing the width of the line to vary as a free parameter. 
In three objects the detection of a broad line is highly significant 
at $>99$\% confidence:  I\,Zw1, PG\,1116+215 and PG\,1402+261. 
 The inferred FWHM are very large with 
64\,000$^{+56\,000}_{-32\,000}$ km\,s$^{-1}$,
  76\,000$^{+51\,000}_{-33\,000}$ km\,s$^{-1}$, 
and  115\,000$^{+61\,000}_{-39\,000}$ km\,s$^{-1}$, respectively.
  In  two objects (PG\,0804+761 and PG\,1309+355)
allowing the line to be broad results in a less significant ($90-99$\%) improvement 
 of the fit statistic. 
 We find smaller FWHM 59\,000$^{+58\,000}_{-23\,000}$ km\,s$^{-1}$,   
 and 23\,000$^{+26\,000}_{-11\,000}$ km\,s$^{-1}$, respectively. 
For PG\,1309+355, the S/N is not sufficient to distinguish 
between a real broad line and a possible blend of a 6.4\,keV line 
with another ionized iron line. 
 For three objects exhibiting both a broad and a highly ionized iron  
line (I\,Zw1, PG\,0804+761, PG\,1116+215), 
the broadening could be due to blend of several ionized iron lines. 
For example, according to the detailed analysis of the data of I\,Zw1,
Gallo et al. (\cite{Ga2004}) proposed that the broad iron emission feature 
can be attributed to a neutral line and a blend of He- and H-like lines. 
However for PG\,1402+261,  the line appears to be blueshifted 
to 7.3 keV (from the broad Gaussian fit) 
beyond the laboratory energy of H-like iron K$_{\alpha}$ fluorescent 
transition, 
and it is therefore unlikely to result from a blend of ionized lines 
between 6.4 keV and 7.0 keV. \\
\indent  For the  five objects showing a broad Fe\,K line, we check 
for possible relativistic effects. 
Therefore, we fit a relativistic line profile:  
 {\sc diskline} from Fabian et al. (\cite{Fabian89}), and 
{\sc laor} from Laor (\cite{La91}) for a non-rotating Schwarzschild BH
 and a rapidly rotating Kerr BH, respectively. 
These  relativistic line profiles give a good fit for all objects. 
For PG 1402+261, the rest-energy of the line is very high (E$\sim$8.5--8.6\,keV)   
when fitted with an AD inclination ($\theta$) fixed at 30 degrees. 
This either implies that the disk is highly inclined to the line-of-sight 
($\theta$=69$^{+8}_{-4}$\,deg, E=6.9$^{+0.3}_{-0.6}$\,keV),  
or that the matter is outflowing with a velocity of about 0.2--0.3\,c assuming 
 a very highly ionized line at 6.7--7\,keV.
 The complex Fe\,K line profile of PG\,1402+261 will be investigated in much 
more detail in a subsequent paper (Reeves et al. 2004, in preparation). 
 No significant difference has been found comparing the 
 $\chi^{2}$ obtained for the relativistic  line profile
and for a broad Gaussian line. The width of the Gaussian lines could be due 
to a blend of several ionized lines 
or simply due to Keplerian motion. 
 With the present S/N data we are not able to infer 
the incidence of real relativistic effects in these quasars. 
Higher S/N data or better spectral resolution 
are needed for such an investigation.\\

\indent  For the five objects showing a significant Fe\,K line consistent
 with 6.4\,keV (PG0947+761, PG\,1048+342, 
PG\,1114+445, PG\,1309+355, PG\,1512+370), 
neither a broad line (as reported in Table~\ref{tab:FeK}), 
nor an additional broad component (i.e. in addition to the narrow line), 
are significantly required (i.e.,  at a confidence level greater than 90$\%$). 
As shown previously by Porquet \& Reeves (\cite{PR2003}), 
broad and intense neutral iron lines in quasars appear to be very rare, 
with the quasar Q0056-363 being the most 
luminous quasar found to date showing such characteristics.  
Interestingly, four of the five NLG in the sample show evidence for  
a highly ionized Fe K line: I\,Zw1, PG\,1115+407, PG\,1244+026, and PG\,1402+261. 
 Five RQQ in our sample show the significant presence of a 
Fe K line, from which only two  correspond to highly ionized iron: 
PG\,0804+761 and PG\,1116+215. These latter two objects have relatively 
steep 2--5\,keV photon indices of about 2.2--2.3. 
Objects with line energies at about 6.4\,keV seem to be  related to
 flatter spectra with $\Gamma\sim$1.4--1.9. 
  The RLQ PG\,1309+355 displays the presence of a ``neutral'' Fe K line, 
whilst the detection of such a line in PG\,1512+379 is less significant. 
Both objects are characterized by a flat photon index of about 1.7. 
 These results are consistent with the correlations found 
between the Fe\,K line energy and 
the 2--10\,keV  X-ray power law slope by Dewangan (\cite{D2002}).
 This  indicates that steep X-ray spectrum objects (such as NLG)
 tend to have highly ionized (e.g. He or H-like) Fe\,K line emission, 
while flatter hard X-ray spectra tend to be associated with neutral
 or weakly ionized material. 
We check this by testing  for Spearman-rank correlations between the line
 energy versus the 2--5\,keV photon index. 
We find a significant positive correlation between both parameters 
with a probability of 99.7$\%$. 
 This correlation between line energy and photon index 
 is consistent with ionized disk reflection models;
 in this scenario the disk iron line emission tends to 
be highly ionized (e.g. He or H-like iron) when the photoionizing continuum is steep, 
and stronger than the emission from 
neutral iron (e.g. Nayakshin et al. \cite{Na2000}, Ross et al. \cite{RFY99}). 
This is due to the formation of a highly ionized layer of iron 
near the accretion disk skin. Therefore one expects to see
strong ionized iron emission lines in NLGs.

\subsection{X-ray disk reflection model}

Given the evidence for broad iron lines in five objects, 
 where the emission is likely to originate from the
inner accretion disk (AD), we attempt to fit the X-ray spectra of these objects  
with a disk reflection model in the 2--12\,keV energy range. 
 We use the ionized disk reflection model {\sc XION} of 
Nayakshin et al. (\cite{Na2000}), in the
most simple configuration where the X-rays are emitted in a "lamppost'' geometry at a
height $h$ above the AD. For this {\sc lamppost} model, we assume  
 a height $h$ for the X-ray source above the disc of 10 R$_{g}$, 
a ratio of X-ray to disk flux of 0.2, 
an outer AD radius of 1000 R$_{g}$, a solar iron abundance,  
 and a high energy cut-off of 100 keV for the emitting source.  
The accretion rate is fixed to the value reported in Table~\ref{tab:mdot}. 
The AD inclination (in deg), and the inner AD radius are free parameters.  
For PG\,1402+261, we need to vary some of the reflection parameters to 
achieve a good fit. We find a lower height for the X-ray source 
above the AD of about 2.5\,R$_{g}$, required to make the emission more 
centrally concentrated around the black hole, whilst 
an iron overabundance of 5 relative to solar is required (the 
maximum the XION model allows) in order to fit 
the extreme strength (EW$\sim2$~keV) of the line. 
Even with this high abundance for PG\,1402+261, 
we find that the spectrum is dominated by the reflected emission from the disc, 
with the hard X-ray power law not directly observed (Reeves et al. 2004, 
in preparation). For the other objects, the spectrum 
is not dominated by the reflection component which  
 contributes less than about 10$\%$ of the total flux at 6 keV.
All  fit results are reported in Table~\ref{tab:lamppost}. 
For all  five objects, this ionized disk reflection model 
gives a good representation of the data over the 2--12\,keV
energy range both for the underlying continuum
 and the Fe\,K$_{\alpha}$ line. \\
\indent In order to check whether the soft excess can also be explained  
by this model, we extrapolate the above fits over the broad 0.3--12\,keV energy range.
Even when leaving the parameters free to vary and adding one or two absorption edges 
 if required, we find a bad representation of the overall spectral shape 
 for the four objects with  a large soft excess. 
Therefore, the soft excess in these objects cannot be solely due to 
reprocessing of the primary X-ray component by an ionized disk, as 
found also for \object{Mrk\,896} (Page et al. \cite{Pa2003}) 
and for Q\,0056-363 (Porquet \& Reeves \cite{PR2003}).
 For I\,Zw1, PG\,0804+761 and PG 1114+445 
which show a weak or no soft excess, the overall spectra are well fitted by an
 ionized disk reflection model.

\section{The broad band spectra: soft X-ray excess and hard tail}\label{sec:all}

\subsection{Thermal emission from the accretion disk?}
The standard explanation for the soft X-ray excess is that it results
  from thermal emission originating directly from the hot inner AD 
(Malkan \& Sargent \cite{MS82}), 
and hence it is the high energy tail of the so-called Big Blue Bump.  
 We fit the data with the combination of an AD black body spectrum 
  ({\sc diskpn} in XSPEC, Gierli{\' n}ski et al. \cite{Gi99})   
 and a power law model, for the soft and hard band, respectively.  
The results are reported in Table~\ref{table:fitbb}. 
 The maximal values of the disk temperature calculated 
 for a standard $\alpha$ thin accretion disk at 3\,R$_{\rm S}$ 
 (Peterson \cite{P97})  are reported in the second column of 
Table~\ref{table:fitbb} (T$^{\rm max}_{\rm disk}$ in eV), 
 using the values of M$_{\rm BH}$ and $\dot{m}$ 
 given in Table~\ref{tab:mdot}. 
The spectra of all  objects are well represented by this model. 
For PG\,1114+445 no disk black body emission is required due to the deep warm absorber 
present in this object.  
 We see that the inferred black body temperatures obtained for the objects 
 appear to be too hot to originate directly from emission of a standard 
$\alpha$ thin accretion disk. 
The only possible exception may be PG\,1244+026 
which has the highest black body to power law flux ratio
below 2\,keV. This quasar has the lowest black hole mass and highest 
Eddington accretion rate 
in the sample, thus one may expect a hotter AD for this object.  

\subsection{Compton scattering?}

As an alternative to the thermal emission model for the soft X-ray excess, 
reprocessing in the form of Compton up-scattering may explain the soft X-ray shape. 
 For example the accretion disk may be responsible for the EUV emission, with some of these 
soft photons being inverse-Compton scattered into the X-ray 
energy range, as they pass through the hot corona above the disk.
Therefore we investigate such models over the 0.3--12 keV range,
by using the {\sc compTT} model in XSPEC (Titarchuk \cite{T94}), 
 using a disk geometry.
We first test a model with one absorbed {\sc compTT} component, 
however we obtain unsatisfactory fits 
 as this model failed to account the soft X-ray curvature observed in most of the spectra.  
Then we assume that there are two Comptonising regions with different temperatures 
and optical depths. 
We fix the input soft photons temperature (kT$_{\rm photon}$)
 of the cooler {\sc compTT} component to the corresponding value of 
  T$^{\rm max}_{\rm disk}$ reported in Table~\ref{table:fitbb} (second column),
 which is representative
of the inner accretion disk temperature. 
 Since our spectral bandpass (up to 12\,keV) does not enable us to 
constrain the temperature of the hotter  {\sc compTT} component, we use a power law model 
which is a good representation for a hotter component 
with a kT$_{\rm plasma}$ of typically 100\,keV
 (Titarchuk \& Lyubarskij \cite{TL95}).
The inferred parameters are reported in Table~\ref{table:fitcomptt}. 
The data are typically fitted with kT$_{\rm plasma}$ 
between 0.1 and 0.4 keV  for the cooler 
 region, except for  PG\,0804+761 and PG\,1114+445 
which exhibit much higher temperature of about 3\,keV.
However the parameters (kT$_{\rm plasma}$ and $\tau$) 
are subject to the inherent degeneracy of the Comptonization models, due to the 
limited bandpass, as pointed out by Brinkmann et al. (\cite{Br2004b}). \\

 In most of the objects in the present sample, 
the soft X-ray excess is too hot to result directly from the inner accretion disk. 
Thus a more realistic physical explanation requires Comptonization of soft (EUV) disk 
photons in a hot plasma. The electron temperature of this soft component 
is typically a few hundred eV, whilst the optical depth 
is high ($\tau>>1$), the result being that the soft X-ray excess resembles a broadened or 
smeared black body component, with an output temperature similar 
to the electron temperature. This 
readily explains the shape of the soft X-ray excess observed in many 
other AGN to date with {\sl XMM-Newton}, 
which can often be fit with multiple black body components 
(e.g. \object{PKS 0558-504}, O'Brien et al. \cite{OB2001}; 
 \object{1H 0419-577}, Page et al. \cite{Pa2002}). 
The power law like component observed above 2 keV is then likely to 
arise from a second much hotter (or even non-thermal) Comptonizing component, where the high 
energy roll-over is expected to be situated outside the {\sl XMM-Newton} bandpass. \\
\indent A plausible physical scenario is that the hard X-ray emission from the 
hot, Comptonizing component heats the surface of the disk, which then emits the cooler 
soft X-ray component that is responsible for the soft excess. 
One problem with this model is the compactness of 
the soft X-ray emission component. The soft Comptonized emission 
can be approximated by a black body as it is  
optically thick. Thus taking typical values from the sample for the soft excess 
temperature ($T\sim10^{6}$~K), soft luminosity ($10^{44}$~erg~s$^{-1}$) and 
black hole mass ($10^{8} \,{\rm M}_{\odot}$), the size of the 
soft X-ray emitting region would only be $R\sim10^{12}$~cm 
or 0.1 gravitational radii ($R_{g}$). Thus 
the soft X-ray component has to be very compact, even for the scenario
 where it is heated by intense 
hard X-ray flares above the disk surface.

\subsection{A relativistically smeared partially ionized absorber?}

Recently, Gierli{\' n}ski \& Done (\cite{Gi2004}) suggested an alternative 
origin for the soft excess  for radio-quiet quasars.  
They proposed that ``the soft X-ray excess is an artefact of strong,
 relativistically smeared, partially ionized absorption''. 
Such absorption requires very steep ($\Gamma>>2$) hard X-ray indices. 
 This would then imply that the hard X-ray continuum emission of 
radio-quiet quasars is different from Seyfert 1 galaxies 
and radio-loud quasars.
 Indeed, Seyfert 1s are constrained from {\sl BeppoSAX} or {\sl RXTE} 
to have hard X-ray indices in the range of $\Gamma=1.6-1.9$ 
(e.g. Perola et al. \cite{Pe2002}), 
as also found in radio-loud quasars, where $\Gamma$ $\sim$ 1.7--1.9 
(Sambruna et al. \cite{Sa2002}). 
For the radio-quiet quasars, this model can only be tested 
 with high S/N data above 10 keV in order to measure
 the hard X-ray photon indices, 
 which will be possible for example with the high sensitivity 
of the hard X-ray detector on the satellite {\sl Astro-E2}.

\section{Global properties and correlations}\label{sec:corr}

We now study the global properties of the present 21 quasars. 
For individual notes on each object, see appendix A.\\

First, we focus on the mean photon index. 
 In the 2--5\,keV energy range, we find an average value 
 of $\Gamma$=1.99  with a standard deviation ($\sigma$) of 0.31 (see Table~\ref{tab:mean}).
We also calculate the mean photon index for each type of quasars present
 in the sample: NLG, RQQ, RLQ. 
The value found for RQQ is consistent with those found from {\sl ASCA} data, 
 i.e. $\Gamma$=1.78$\pm$0.11 ($\sigma$=0.29$^{+0.09}_{-0.07}$), 
and $\Gamma$=1.89$\pm$0.05 
(Leighly \cite{L99} and  Reeves \& Turner \cite{RT2000}, respectively).  
NLG have the steepest power law index compared to optical broad line quasars, 
such as RQQ and RLQ  (Table~\ref{tab:mean}). 
This confirms previous results and is compatible with the photon index 
found with {\sl ASCA}: 2.19$\pm$0.10 ($\sigma$=0.30$^{+0.07}_{-0.06}$), 
and 2.12 ($\sigma$=0.26), by Leighly (\cite{L99}) 
and Vaughan et al. (\cite{V99}), respectively.
 The mean soft photon index  (0.3--2\,keV)  is high with  
2.56 (with a standard deviation of 0.44), consistent with  {\sl ROSAT} results
 ($\Gamma$= 2.55$\pm$0.33, Wang et al. \cite{WBB96}; 
$\Gamma$= 2.58$\pm$0.05 Yuan et al. \cite{Y98}). 
For NLG the mean soft photon index is even higher with 3.04  ($\sigma$=0.42) 
and consistent with the mean values of 3.29$\pm$0.61 
calculated by Wang et al. (\cite{WBB96}) for  objects
 exhibiting extreme \ion{Fe}{ii} emission 
 (including I\,Zw1, PG\,1244+051, PG\,1402+261, Mrk\,478). 
The values reported in Table~\ref{tab:mean} strongly suggest 
 that the steepest soft X-ray spectra correspond 
 to the steepest hard X-ray spectra. This is confirmed below. \\

\begin{figure*}[t!]
\begin{tabular}{@{}c@{}c@{}c}
 \psfig{file=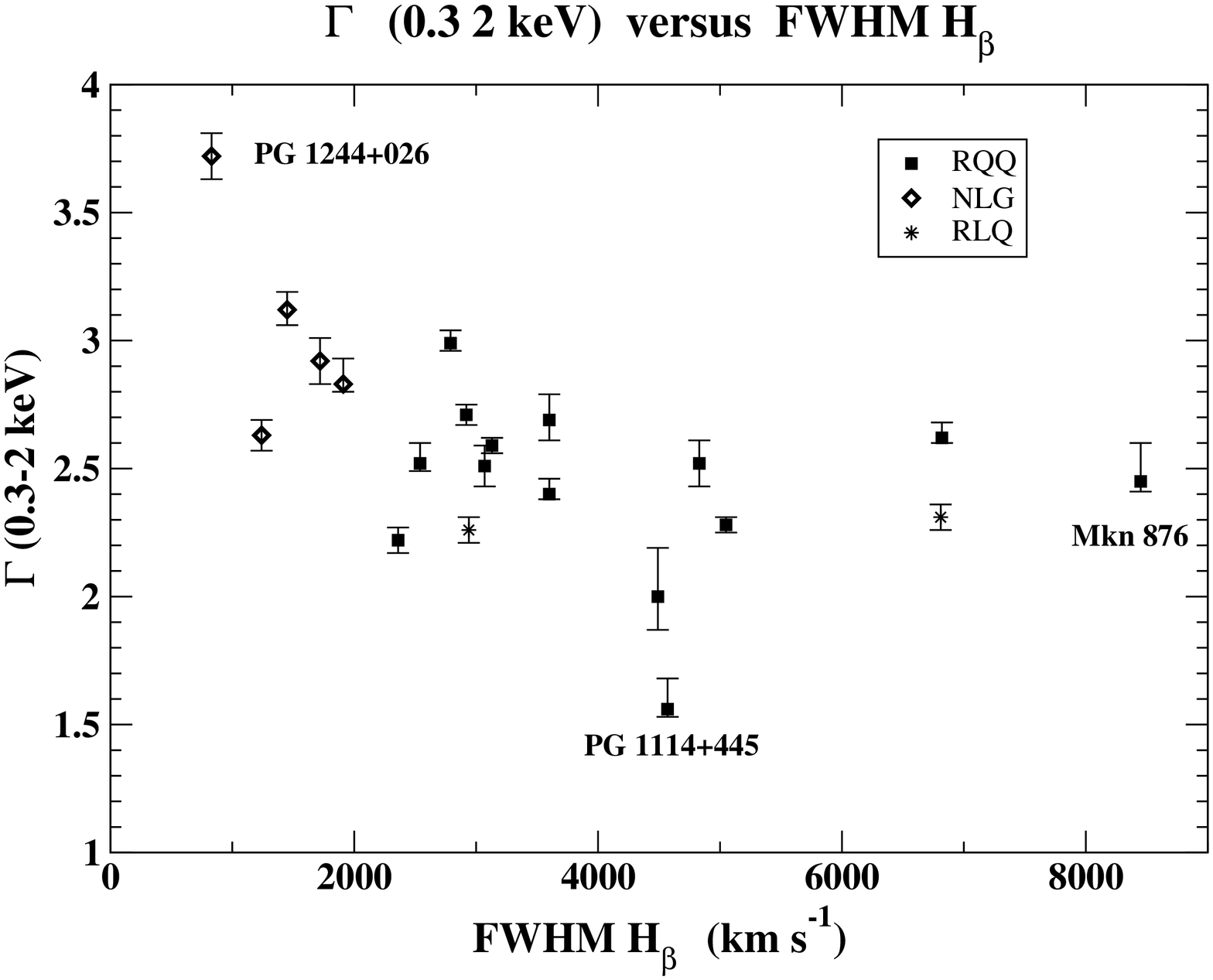,width=6.cm,angle=-90} &\psfig{file=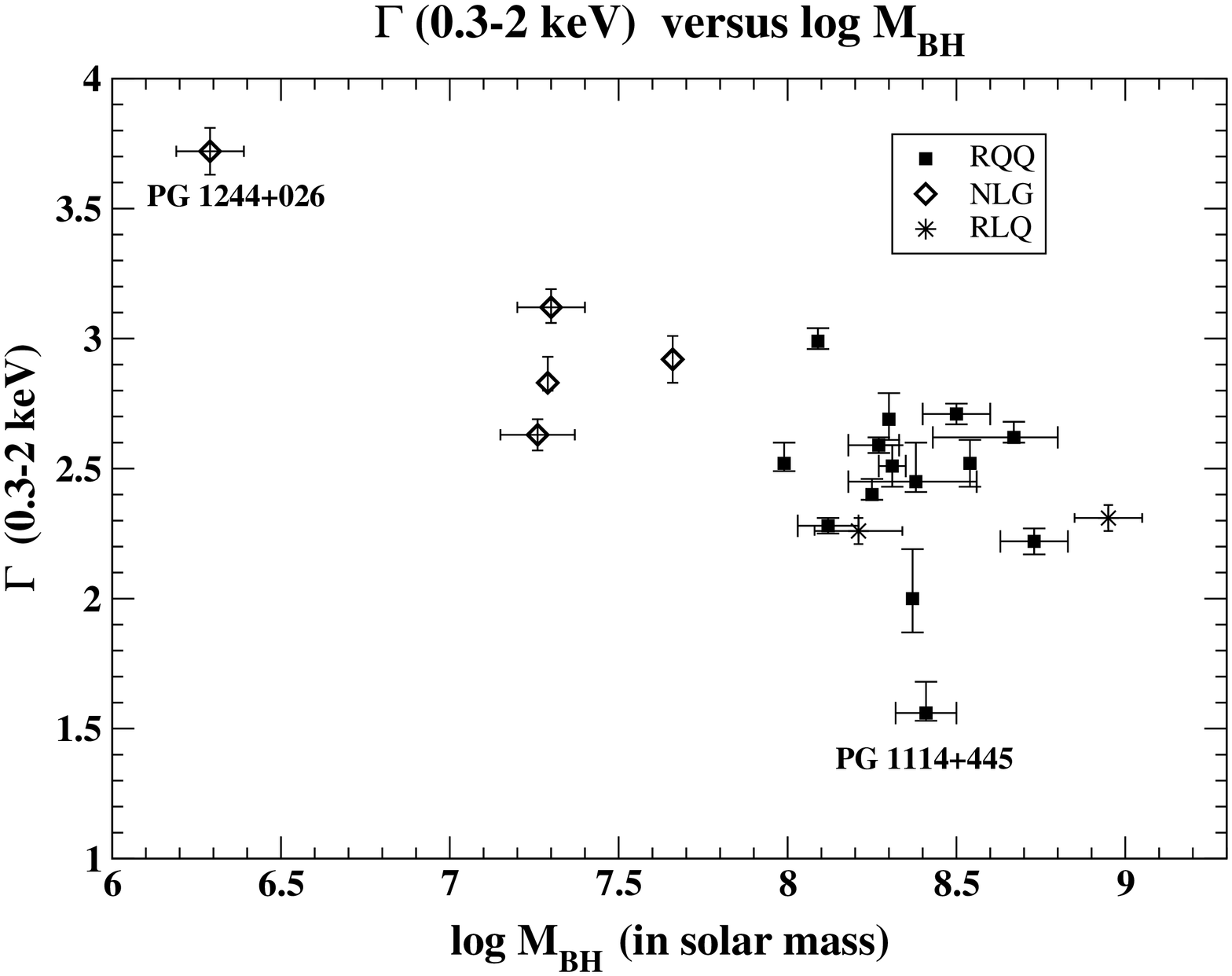,width=6.cm,angle=-90} &  \psfig{file=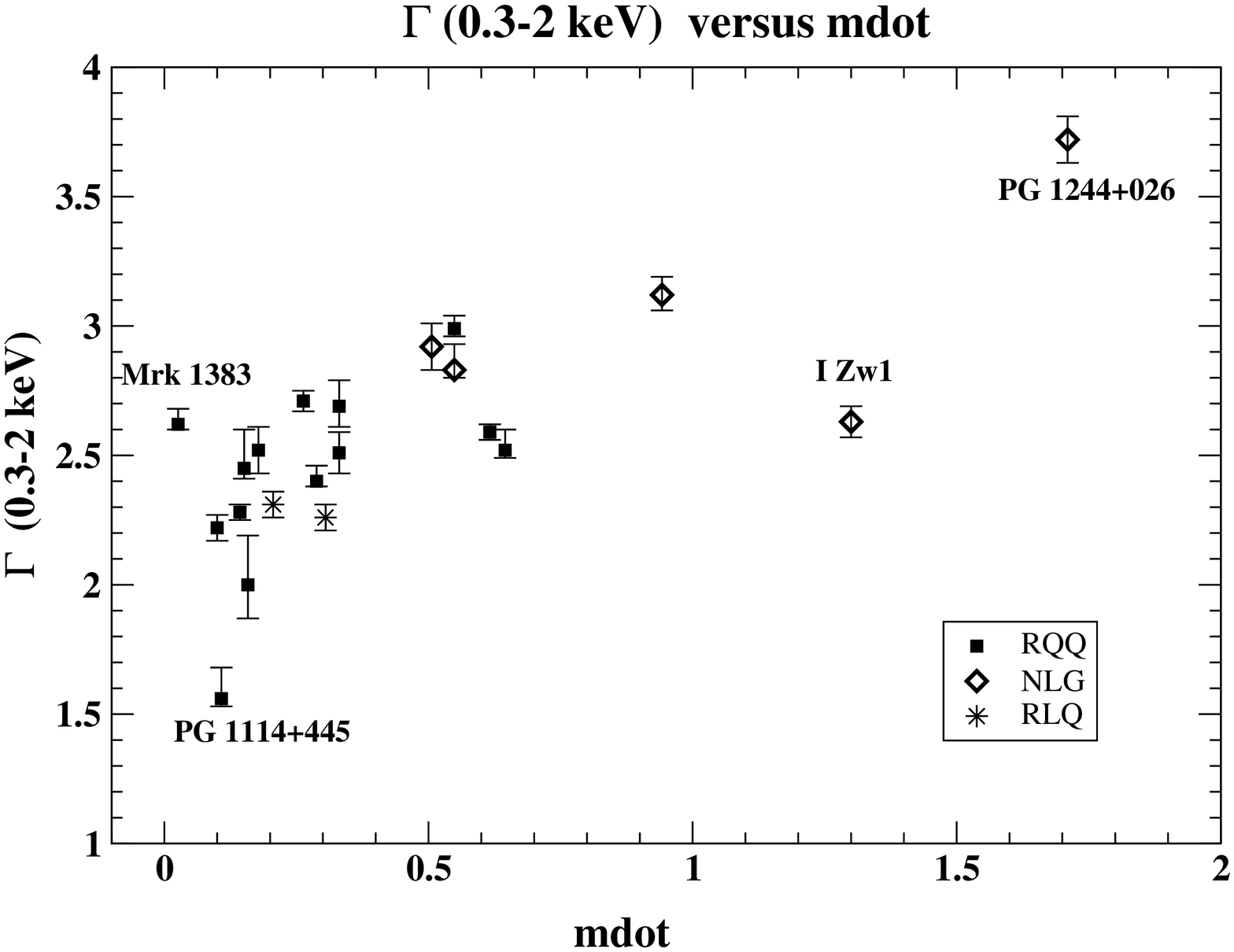,width=6.cm,angle=-90} \\
 \psfig{file=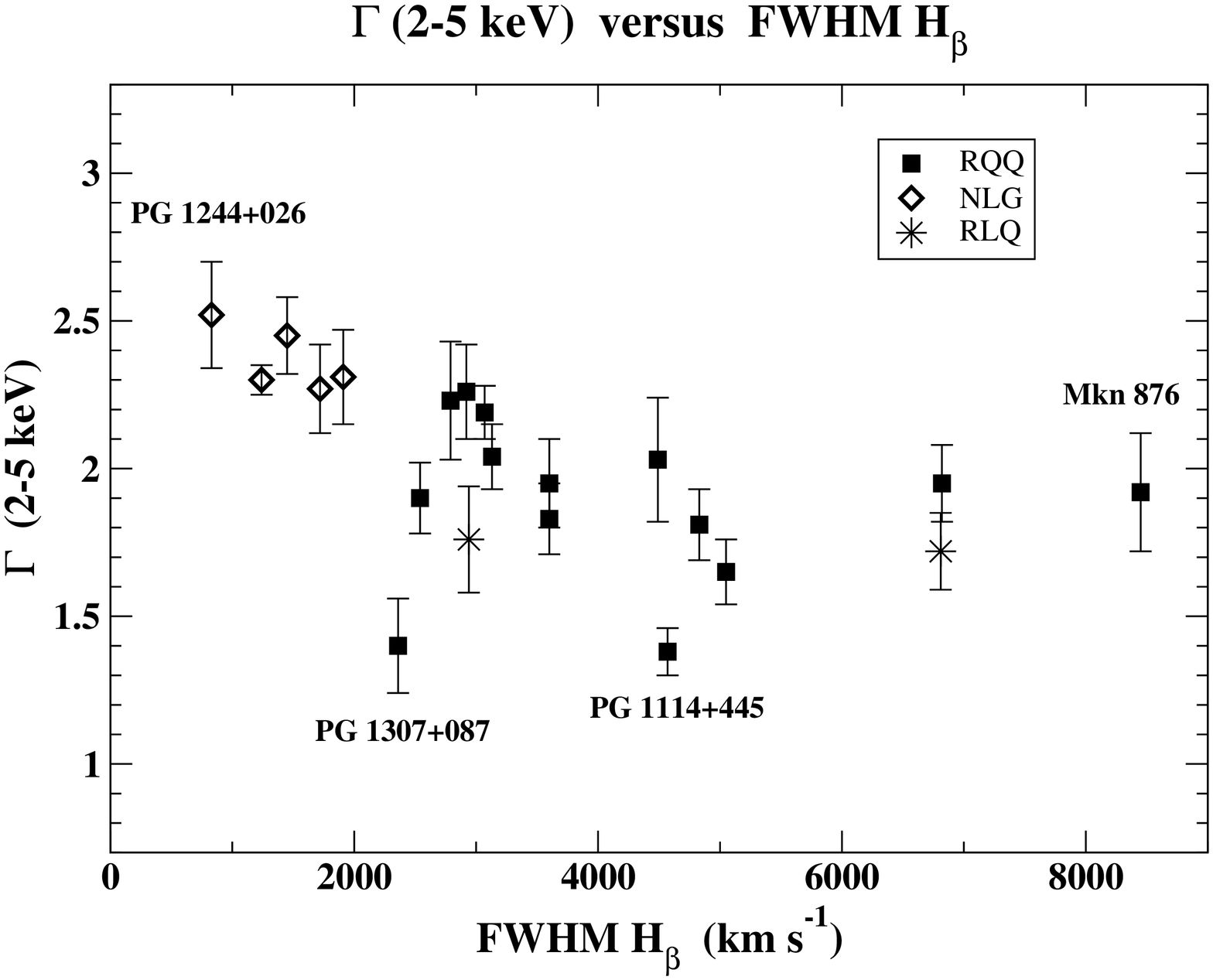,width=6.cm,angle=-90}  & \psfig{file=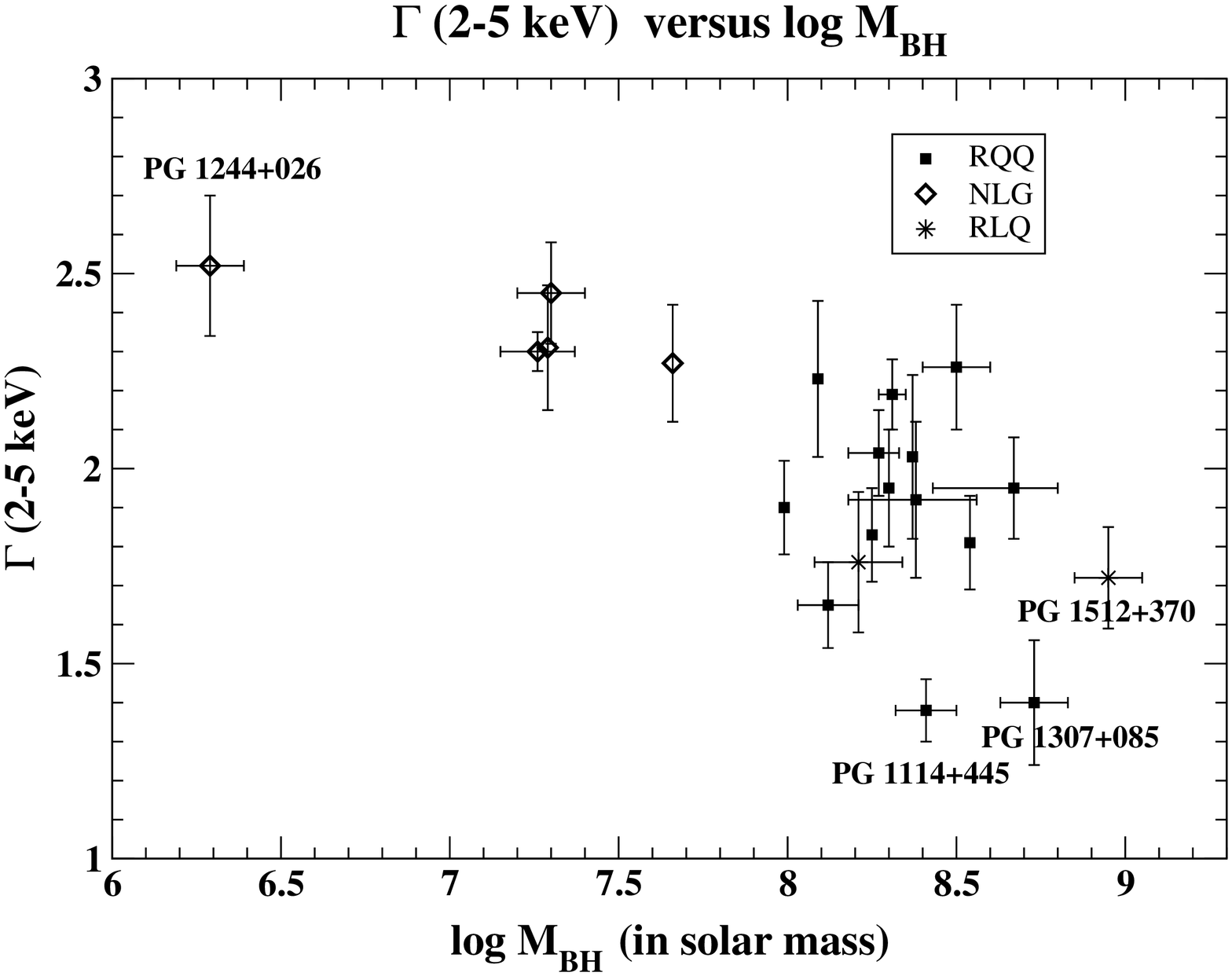,width=6.cm,angle=-90} & \psfig{file=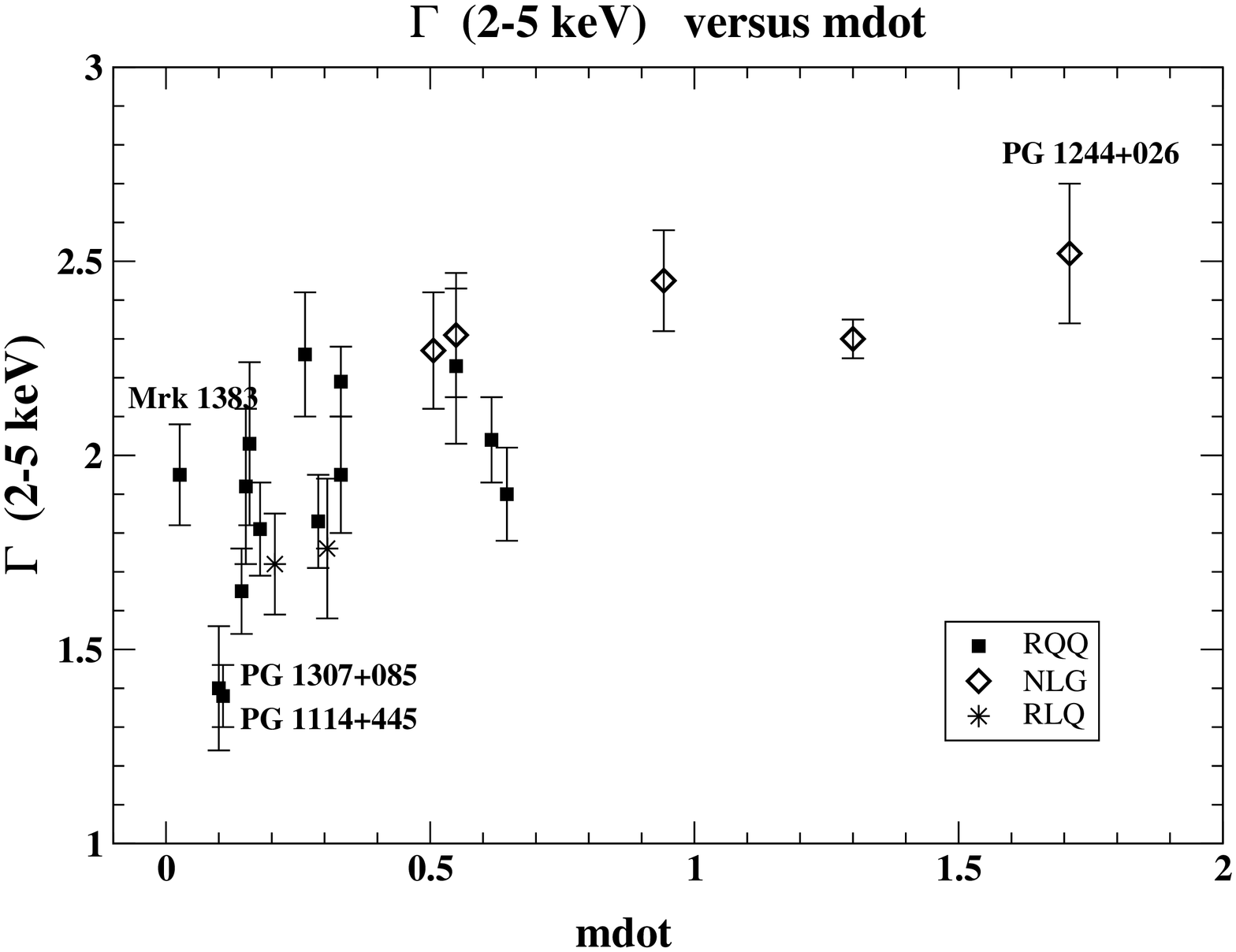,width=6.cm,angle=-90}\\
 \psfig{file=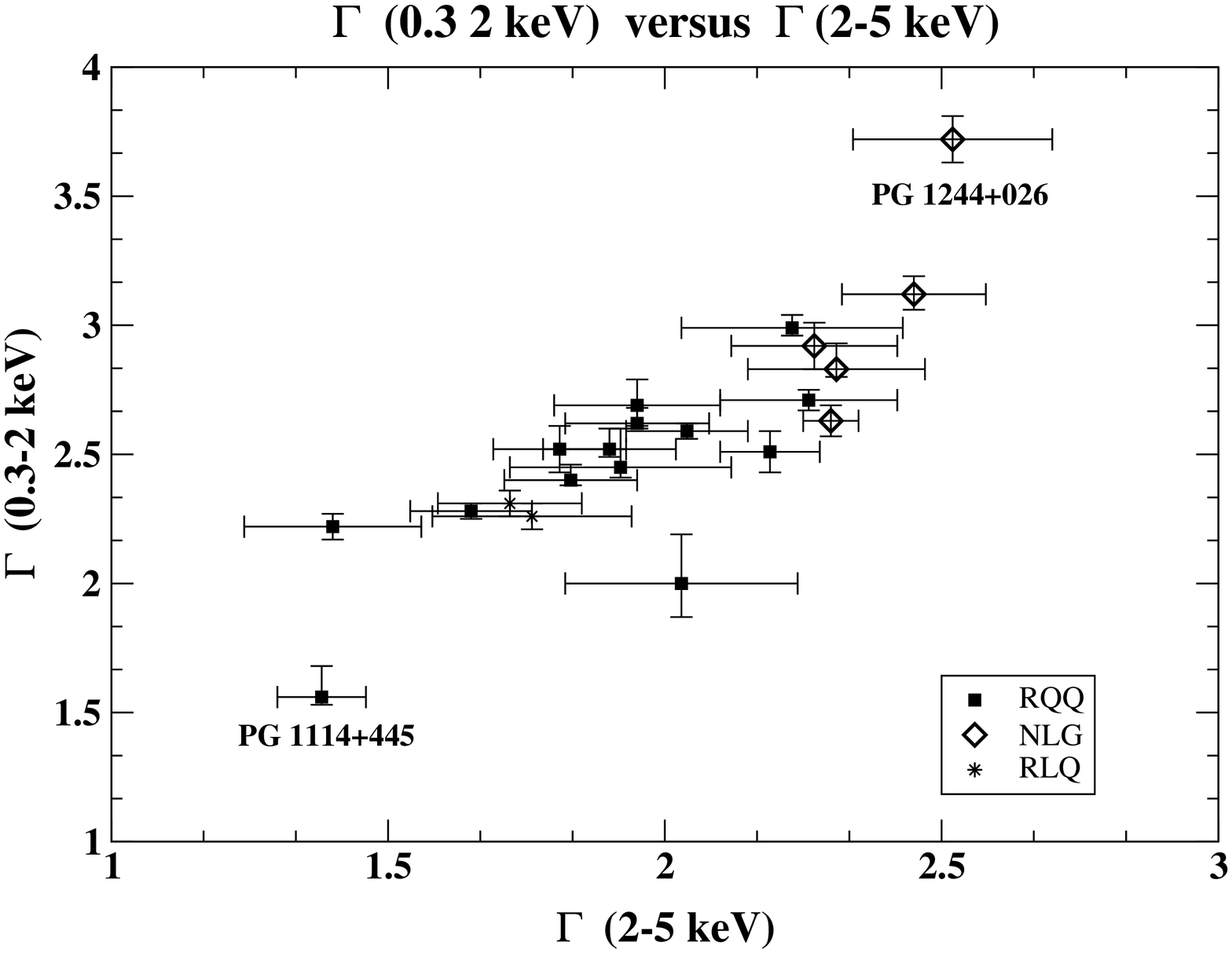,width=6.cm,angle=-90}\\ 
\end{tabular}
\caption{Correlations between X-ray and AGN properties.}
\label{fig:corr}
\end{figure*}

We now test some Spearman-rank correlations between the X-ray parameters 
($\Gamma_{\rm 0.3-2}$, $\Gamma_{\rm 2-5}$,  L$_{0.3-2}$, L$_{2-10}$), 
and some quasar characteristics: FWHM(H${\beta}$), radio-loudness (R$_{\rm L}$), 
the black hole mass (M$_{\rm BH}$), and the accretion rate ($\dot{m}$). 
The results are reported in Table~\ref{tab:corr}. 
We find a very strong anti-correlation between 
$\Gamma_{\rm 2-5}$, $\Gamma_{\rm 0.3-2}$ and the FWHM(H$\beta$) line 
(Fig.~\ref{fig:corr}). 
This confirms the results found previously 
(e.g., Boller et al. \cite{Bo96}, Wang et al. \cite{WBB96}, 
Brandt et al. \cite{B97}, Vaughan et al. \cite{V99}, 
Reeves \& Turner \cite{RT2000}, and Dewangan \cite{D2002}). 
Indeed spectra appear much softer in objects with narrow permitted optical H$\beta$ lines. 
We find also a very significant correlation between $\Gamma_{\rm 0.3-2}$ 
and $\Gamma_{\rm 2-5}$ (Fig.~\ref{fig:corr}), 
meaning that a steep soft photon index leads to a steep hard photon index, consistent with 
Compton cooling models. 
Both, soft and hard photon index are strongly anti-correlated with the black hole mass, 
and positively correlated with the accretion rate relative to
 Eddington (Fig.~\ref{fig:corr}).
Thus the steepest objects have the lowest black hole masses and the highest accretion rates.
 Therefore, a high accretion rate and a smaller black hole mass is
 likely to be the physical driver 
responsible for these trends, leading to a higher 
inner disk temperature, which results in a stronger soft X-ray excess
 (and therefore a steeper soft photon index) and subsequently a steep hard power law.

\section{Summary and Conclusions}

We have presented results from {\sl XMM-Newton} EPIC observations of 21 low 
redshift quasars with redshifts less than 0.4.  All  objects are Palomar Green 
quasars and have low Galactic absorption along the line-of-sight. 
Five objects are Narrow Line Seyfert1/quasars, two are RLQ, and the other
 14 objects are RQQ. We now summarize our main results: \\

About 90$\%$ of the present quasars (19/21) 
exhibit a significant soft excess at energies below about 1--1.5\,keV. 
This result is in contrast to previous findings with {\sl ASCA}, e.g. 
George et al. (2000) found that only five of their 14 PG quasars  
(with z$\leq$0.25) require a soft excess below 1\,keV. This is 
likely due to the smaller band-pass of {\sl ASCA} compared 
to {\sl XMM-Newton} at lower energies. 
For objects with a strong soft excess and a large Fe\,K line, the soft excess 
 cannot be solely due to reprocessing of the primary X-ray component 
by an ionized disk. 
In all objects in the present sample (except possibly PG\,1244+026), 
the soft X-ray excess is much better explained by Comptonization 
of soft photons in a hot plasma,
 than by the thermal emission from a thin standard accretion disk, 
as the inferred disc temperatures appear to be too hot. 
However, to definitely discriminate between these two models, 
a larger bandpass is required, from the 
EUV (for the soft component) to hard X-rays up to few 100\,keV
 (for the hard component).\\

PG\,1114+445 shows a huge absorption trough between 0.6 and 1.5\,keV, 
due to the presence of a warm absorber. 
Fitting the spectrum with two absorption edges, we find very 
large optical depths of 
  $\tau\sim$2.23 and  $\tau\sim$0.53  
at E=0.72\,keV and E=0.93\,keV, respectively.
 Further nine other objects show the presence 
of one or two absorption edges near those of \ion{O}{vii}
 (0.74\,keV) and \ion{O}{viii} (0.87\,keV).
This means that at least 50$\%$ of our present quasars show absorption,
 which is compatible with the fraction found in Seyfert\,1 galaxies
 at lower redshift (e.g., Reynolds \cite{Re97} with {\sl ASCA}).
 This is the first time that such a high fraction of warm absorbers is 
found in quasars and may imply a similar covering fraction of the 
absorbing material compared to the Seyfert 1s. 
The lack of detection of warm absorbers in quasars with {\sl ROSAT} 
and {\sl ASCA} may have been due to the limited bandpass for 
both instruments, i.e. the lack of soft response of {\sl ASCA} and the limited 
spectral resolution of the {\sl ROSAT} PSPC. 
The present data do not enable us to determine the exact
nature (ionization degree, velocity, etc.)  and the complexity  
 (e.g., absorption/emission lines, absorption edges) 
of the absorbing medium. Higher spectral resolution data are required (e.g. RGS),
 which will be investigated in a subsequent paper. \\

We find evidence for Fe K shell emission lines in at least twelve objects.
 Indeed, this is probably a lower limit on the number of 
objects with iron lines, since the line detection depends on the exposure times
 which are very short in some objects. 
Nonetheless, iron lines are detected in five objects at $>$99$\%$ confidence
 (I\,Zw1, PG\,0804+761, PG\,1114+445, PG\,1116+215 and PG\,1402+261), 
whilst seven objects have a more marginal detection at 90--99$\%$ confidence
 (PG\,0947+396, PG\,1048+342, PG\,1115+407, PG\,1244+026, 
PG\,1309+355, Mkn\,1383, and PG\,1512+370).
   In five objects the line energy is consistent within the errors with 6.4 keV, 
 a ``neutral'' to moderately ionized iron (i.e. $\leq$ \ion{Fe}{XVII}), while  
 for the other seven objects the line energy is consistent 
with  highly ionized iron:  I\,Zw1, PG\,0804+761, PG\,1115+407, PG\,1116+215,
 PG\,1244+026, PG\,1402+261, and Mrk\,1383. 
For  five objects the fit is statistically improved when leaving the
 width of the line as a free parameter. 
 All these line profiles can be well fitted with either a broad Gaussian line, 
or with a relativistic line profile (non-rotating Schwarzschild BH 
or rapidly rotating Kerr BH). 
 No significant differences have been found comparing the 
 $\chi^{2}$ obtained for the relativistic line profile 
and for the broad Gaussian line. The width of the Gaussian line could be due 
to a blend of several (neutral/ionized) lines or simply to Keplerian motion. 
 With the present S/N data we are not able to infer the incidence
 of genuine relativistic effects in quasars. 
Higher S/N data or larger spectral resolution are needed
 for such investigation. 
The line profiles can be also be interpreted as ionized disk reflection, 
which is able to explain the spectral shape over the 2--12\,keV energy range. 
At least five significant detections (at $\geq$ 99$\%$ confidence) of 
broad iron lines in quasars have been found with {\sl XMM-Newton} to date: 
I\,Zw1, PG\,0804+761, PG\,1116+215, 
 and PG\,1402+261 (this work), and Q\,0056+363 (Porquet et al. \cite{PR2003}). 
 The first three objects exhibit line energies corresponding 
to very ionized iron (He-like, H-like).
 As shown previously by Porquet \& Reeves (\cite{PR2003}), 
broad and intense lines at 6.4\,keV in quasars appear to be very rare, 
with the quasar Q0056-363 being the most
 luminous quasar found to date exhibiting such characteristics.  
 
The RQQ PG\,1402+261 shows a very large positive deviation
 near 7--8\,keV compared to a power law model.  
 Assuming that this feature is a relativistic Fe\,K line, 
we find a huge EW of about 2\,keV, 
thus the hard X-ray spectrum of this object is likely to be 
dominated by reflected emission from the disc. Other interpretations 
such as partial covering absorber models will be investigated in detail in a \
forthcoming paper (Reeves et al. 2004, in preparation).
 Four of the five NLG of the sample show the presence of 
a very highly ionized  Fe\,K line: I\,Zw1, PG\,1115+407, 
PG\,1244+026, and PG\,1402+261. 
 Our results are also consistent with the correlations found between the
 Fe\,K line energy and 
the 2--10\,keV  X-ray power law slope by Dewangan (\cite{D2002}). This means 
that the steep X-ray spectrum objects (such as NLG) tend to have Fe\,K 
lines formed in a highly ionized medium, 
while objects with flatter X-ray spectra tend to be associated with 
near neutral or weakly ionized iron line emission. \\

A strong correlation is found between
 $\Gamma$ (both in the soft 0.3--2\,keV and hard 2--10\,keV energy bands)  
and optical H$\beta$ width, whereby the steepest X-ray spectra
tend to be found in those objects with narrow H$\beta$ widths. 
This is consistent with previous results 
 (e.g., Boller et al. \cite{Bo96}, Wang et al. \cite{WBB96}, 
Brandt et al. \cite{B97}, Vaughan et al. \cite{V99}, 
Reeves \& Turner \cite{RT2000}, and Dewangan \cite{D2002}). 
The soft and hard X-ray photon indices are also linked by a very strong 
correlation, 
i.e. the steepest soft X-ray spectra lead to the steepest hard X-ray spectra. 
 The strongest correlations are found between $\Gamma$ (soft and hard), 
H$\beta$ width  and the black hole mass and the accretion rate. 
 Therefore, we conclude that a high accretion rate and a smaller black hole mass
 is likely to be the physical driver responsible for these trends, 
i.e. the steepest X-ray spectra are often 
found in objects accreting at high fraction of the Eddington rate, 
with smaller black hole masses. \\

\section*{Acknowledgments}
Based on observations obtained with the {\sl XMM-Newton}, an ESA science
mission with instruments and contributions directly funded by ESA
member states and the USA (NASA). We would like to thank the anonymous
referee for fruitful comments and suggestions. 
D.P. acknowledges grant support from an MPE fellowship.

\clearpage
\clearpage

\clearpage
\begin{table*}[ht!]
\caption{The 21 Type I AGN included in the present sample. 
The Galactic column density along the line-of-sight (N$_{\rm H}^{\rm Gal}$) is expressed  in units of 10$^{20}$\,cm$^{-2}$. 
NLG: narrow line galaxy. RQQ: radio-quiet quasar. RLQ: radio-loud quasar. 
}
\begin{tabular}{llllllllcl}
\hline
\hline
\noalign {\smallskip}                       
\multicolumn{1}{c} {Source}  &  \multicolumn{1}{c} {other name}    &\multicolumn{2}{c} {Coordinates (J2000)} & \multicolumn{1}{c} {z} &  N$^{\rm Gal}_{\rm H}$ & \multicolumn{1}{c} {M${\rm v}$}  &  \multicolumn{1}{c} {R$_{\rm L}$}   &  \multicolumn{1}{c} {FWHM(H$_{\beta}$)}  &\multicolumn{1}{c} {Type}\\
                          &                         & \multicolumn{1}{c} {RA} & \multicolumn{1}{c} {Dec }      &   &                              &                &                            & km\,s$^{-1}$    \\
\hline
\noalign {\smallskip}  
\object{PG\,0050+124}   &   \object{I\,Zw1} & 00 53 35.08     & +12 41 34.4                 & 0.061 &   4.94     &  $-$23.77    &     $-$0.48         &  1240 & NLG       \\
\object{PG\,0804+761}   &                         &  08 10 58.66 & +76 02 42.5                  & 0.100  &   3.01        & $-$24.44      & $-$0.22  &3070  &RQQ \\
\object{PG\,0947+396}   &  \object{K347-45}          &  09 50 48.39  & +39 26 50.6                 & 0.206  & 1.55  & $-$24.21      & $-$0.60  &4830 &RQQ \\ 
\object{PG\,0953+414}   &  \object{K348-7}          & 09 56 52.41   &  +41 15 22.1               & 0.239      & 0.82   & $-$25.65     & $-$0.36  &  3130&RQQ \\ 
\object{PG\,1048+342}  &                          &  10 51 43.88  & +33 59 26.9                  & 0.167  &  1.94  & $-$24.02        &  $-$1.00   &  3600 &RQQ \\ 
\object{PG\,1114+445}   &                      &     11 17 06.40 & +44 13 33.2                   & 0.144 &  1.62  & $-$24.01       & $-$0.89    & 4570 &RQQ\\
\object{PG\,1115+407}  &                      &  11 18 30.30 &  +40 25 53.8 & 0.154  &1.91 &$-$23.74      & $-$0.77  & 1720  & NLG    \\ 
\object{PG\,1116+215}   &  \object{TON 1388} & 11 19 08.60 & +21 19 18.0 & 0.177 &1.29 & $-$25.57      & $-$0.14  & 2920  &RQQ    \\
\object{PG\,1202+281}   &   \object{GC Com}       &     12 04 42.11 &  +27 54 11.5 & 0.165& 1.70  &  $-$23.75  &  $-$0.72    &5050&RQQ \\
\object{PG\,1244+026}   &                      &     12 46 35.29   & +02 22 08.3   & 0.048 &1.78  & $-$21.77& $-$0.28 & 830  &NLG \\ 
\object{PG\,1307+085}   &                      &    13 09 47.03  & +08 19 49.3  &0.155 & 2.10  & $-$24.56      & $-$1.00  & 2360 &RQQ \\ 
\object{PG\,1309+355}   &  \object{TON 1565}   &    13 12 17.74   & +35 15 21.3  & 0.184 &1.02 & $-$24.76     & $+$1.26  & 2940  &RLQ \\ 
\object{PG\,1322+659}   &                                &   13 23 49.54  &  +65 41 48.0 & 0.168 &2.00 & $-$24.23     & $-$0.92  & 2790  &RQQ\\ 
\object{PG\,1352+183}      & \object{PB 4142}   & 13 54  35.93 & +18 05 18.1  & 0.152 &  2.05         & $-$24.13    & $-$0.96  & 3600 &RQQ\\
\object{PG\,1402+261}  &      \object{TON 182}   &   14 05 16.21  & +25 55 34.1  & 0.164 &1.47  &  $-$24.48   &  $-$0.64  & 1910 & NLG\\ 
\object{PG\,1426+015}    & \object{Mrk 1383}    &  14 29 06.59 &  +01 17 06.5 & 0.086       & 2.85  &  $-$24.05  & $-$0.55  & 6820 &RQQ\\
\object{PG\,1427+480}   &                      & 14 29 43.08   &+47 47 26.3 &  0.221  &  1.86   &  $-$24.04  & $-$0.80 &  2540  &RQQ \\ 
\object{PG\,1440+356}   &  \object{Mrk 478}      & 14 42 07.46     & +35 26 22.9     & 0.077   &  1.04 & $-$23.49  &  $-$0.43   & 1450  &NLG        \\
\object{PG\,1512+370}    &  \object{4C 37.43}     &15 14 43.06  &+36 50 50.5 &  0.371  & 1.35 & $-$25.93      & $+$2.28 & 6810 &RLQ\\
\object{PG\,1613+658}   &  \object{Mrk 876}      & 16 13 57.18 & +65 43 09.6 & 0.129   & 2.85 & $-$24.22     & $+$0.00    & 8450  &RQQ  \\
\object{PG\,1626+554}  &                      &  16 27 56.09  & +55 22 31.7 & 0.132 &2.02 & $-$23.54    & $-$0.96  & 4490 &RQQ   \\ 
\noalign {\smallskip}                       
\hline
\hline
\end{tabular}
\label{tab:sample}
\end{table*}
\clearpage

\begin{table}[h!]
\caption{List of the {\sl XMM-Newton} observations. Column (1): Name of the source; 
 (2): observation date (dd/mm/yyyy),  (3): orbit revolution number, 
 (4): Observation ID, and (5): data exposure time of the pn data (in ks), 
 except for PG\,0804+761 for which the MOS exposure time is given.}
\begin{center}
\begin{tabular}{llllr}
\hline
\hline
\noalign {\smallskip} 
Source               &   \multicolumn{1}{c}{Obs. Date} &  \multicolumn{1}{c}{Rev.} & \multicolumn{1}{c}{Obs-ID}     &\multicolumn{1}{c} {Exp. }  \\
\noalign {\smallskip} 
\hline
\noalign {\smallskip} 
I\,Zw1 &    22/06/2002  & 464    &  0110890301  & 18.7\\
PG\,0804+761 &    04/04/2000  &  166&0102040401 & 6.7\\
 PG\,0947+396  &   03/11/2001& 349&0111290101 &17.6 \\ 
PG\,0953+414  &    22/11/2001& 358&0111290201&11.5 \\
PG\,1048+342  &    13/05/2002 &444&0109080701 & 28.1  \\ 
PG\,1114+445  &  14/05/2002  & 445& 0109080801& 35.0 \\
PG\,1115+407 &  02/12/2001 &446& 0111290301 & 15.0  \\ 
PG\,1116+215  & 02/12/2001 & 363& 0111290401& 5.5  \\
PG\,1202+281  &  30/05/2002 &453& 0109080101&12.0  \\
PG\,1244+026  &  17/06/2001 &279& 0051760101&7.2  \\
PG\,1307+085  &  13/06/2002   &460& 0110950401 & 11.2  \\ 
PG\,1309+355  &  10/06/2002 & 458& 0109080201  & 22.3  \\ 
PG\,1322+659  & 11/05/2002 & 443& 0109080301& 8.7  \\ 
PG\,1352+183  &  20/07/2002 & 478 & 0109080401 & 12.5\\
PG\,1402+261  &   27/02/2002 &391& 0109081001& 9.1  \\
Mrk 1383   &  28/07/2000& 116  & 0102040501 & 5.4\\
PG\,1427+480  & 31/05/2002 & 453& 0109080901 & 31.2  \\
Mrk 478 & 23/12/2001  & 373       & 0107660201  &9.1\\             
PG\,1512+370   & 25/08/2002 & 496 &   0111291001  &  17.6   \\             
Mrk 876              & 29/08/2001 & 315  &  0102041301  & 2.5 \\
PG\,1626+554   & 05/05/2002 &440& 0109081101 & 5.5  \\
\noalign {\smallskip}                       
\hline
\hline
\end{tabular}
\end{center}
\label{tab:obs}
\end{table}
\begin{table}[h!]
\small
\caption{Black hole mass (M$_{\rm BH}$), 
 monochromatic luminosity  $\nu$L$_{\nu}$(5100\AA),   
 bolometric luminosity (L$_{\rm bol}$),
 and accretion rate with respect to Eddington.
(*) L$_{\rm bol}\simeq$ 9 $\times$ $\nu$L$_{\nu}$(5100\AA). 
(**) using the relation (3) given by Woo \& Urry et al. (\cite{WU2002}) between M$_{\rm BH}$ 
and $\nu$L$_{\nu}$(5100\AA). 
(a): Kaspi et al. (\cite{K2000}),  (b): Vestergaard (\cite{Ves2002}), 
and (c): Shields et al. (\cite{Sh2003}), (d): Grupe et al. (\cite{G2004}), 
and (e):  Woo \& Urry (\cite{WU2002}).   
}
\vspace*{-0.3cm}
\begin{center}
\begin{tabular}{@{}ll@{}l@{}l@{}l}
\hline
\hline
\noalign {\smallskip}  
\multicolumn{1}{c}{Source } & \multicolumn{1}{c}{log\,M$_{\rm BH}$} &   \multicolumn{1}{c}{{\small log\,$\nu$L$_{\nu}${\tiny(5100\AA)}} }  &  \multicolumn{1}{c}{log\,L$_{\rm bol}$}   &  \multicolumn{1}{c}{$\dot{m}$} \\
                            &\multicolumn{1}{c}{(M$_{\odot}$)} &  \multicolumn{1}{c}{ (erg\,s$^{-1}$)}& \multicolumn{1}{c}{(erg\,s$^{-1}$)}   &   \\
\noalign {\smallskip}  
\hline
\noalign {\smallskip}  
I\,Zw1  & 7.26$\pm$0.11$^{(b)}$       &   44.52$\pm$0.12$^{(b)}$    &    45.47$\pm$0.12$^{(*)}$       &  1.30\\
PG\,0804+761&  8.31$\pm$0.04$^{(a)}$       &  44.82$^{+0.07}_{-0.09}$$^{(a)}$  & 45.93$^{(e)}$                     & 0.33\\
PG\,0947+396& 8.54$^{(c)}$                    & 44.94$^{(c)}$                 & 45.89$^{(*)}$                   & 0.18            \\ 
PG\,0953+414 & 8.27$^{+0.06}_{-0.09}$$^{(a)}$ & 45.08$^{+0.05}_{-0.07}$$^{(a)}$ & 46.16$^{(e)}$  & 0.62 \\
PG\,1048+342 & 8.25$^{(c)}$                   & 44.86$^{(c)}$                   & 45.81$^{(*)}$                   & 0.29      \\ 
PG\,1114+445 & 8.41$\pm$0.09$^{(b)}$   & 44.59$\pm$0.07$^{(b)}$   &  45.54$^{(*)}$           & 0.11\\
PG\,1115+407 &     7.66$^{(**)}$                  &     44.51$^{(d)}$                & 45.83$^{(d)}$                  & 0.51      \\ 
PG\,1116+215 &   8.50$\pm$0.10$^{(b)}$       &  45.27$\pm$0.07$^{(b)}$           & 46.02$^{(e)}$   & 0.26\\
PG\,1202+281 & 8.12$\pm$0.09$^{(b)}$         &  44.42$\pm$0.07 $^{(b)}$        & 45.26$^{(d)}$         &0.14\\ 
PG\,1244+026 & 6.29$\pm$0.10$^{(b)}$   &  43.67$\pm$0.08$^{(b)}$        & 44.75$^{(d)}$         & 1.71\\ 
PG\,1307+085  &8.73$\pm$0.10$^{(b)}$ & 44.86$\pm$0.07$^{(b)}$          & 45.83$^{(e)}$  & 0.10\\
PG\,1309+355  &8.21$\pm$0.13$^{(b)}$         & 44.84$\pm$0.07$^{(b)}$          & 45.82$\pm$0.07$^{(*)}$          & 0.31\\ 
PG\,1322+659  &  8.09$^{(c)}$                 & 44.98$^{(c)}$                    & 45.93$^{(*)}$                   & 0.55 \\ 
PG\,1352+183  &    8.30$^{(c)}$           & 44.97$^{(c)}$                     & 45.92$^{(*)}$                   &     0.33\\
PG\,1402+261               & 7.29$^{(e)}$               &                                &  45.13$^{(e)}$   & 0.55\\
Mrk\,1383  &8.67$^{+0.13}_{-0.24}$$^{(a)}$ & 44.61$^{+0.06}_{-0.07}$$^{(a)}$  & 45.19$^{(e)}$  & 0.03 \\
PG\,1427+480  & 7.99$^{(c)}$                 &  44.95$^{(c)}$                    &  45.90$^{(*)}$                  & 0.65  \\ 
Mrk\,478     &   7.30$\pm$0.10$^{(b)}$     &   44.42$\pm$0.07$^{(b)}$        &  45.88$^{(d)}$        & 0.94\\
PG\,1512+370  &   8.95$\pm$0.10$^{(b)}$      & 45.41$\pm$0.06$^{(b)}$        &  46.36$\pm$0.06$^{(*)}$         &0.21\\
Mrk\,876   & 8.38$^{+0.18}_{-0.20}$$^{(a)}$                  &      44.84$\pm$0.06$^{(a)}$  &  45.66$^{(e)}$         &   0.15 \\
PG\,1626+554  &  8.37$^{(c)}$                 &  44.72$^{(c)}$                   &   45.67$^{(d)}$                 &   0.16\\ 
\noalign {\smallskip}                       
\hline
\hline
\end{tabular}
\end{center}
\label{tab:mdot}
\end{table}

\clearpage
\begin{table*}[h!]
\caption{Spectral fits with an absorbed power law continuum in two energy bands
 (2--5\,keV and 0.3--2\,keV). 
Galactic absorption (N$_{\rm H}^{\rm Gal}$) 
 and intrinsic absorption (N$^{\rm in}_{\rm H}$) of the QSO in its rest frame
 are expressed in 10$^{20}$\,cm$^{-1}$.
 (a): Absorption edge(s) added to the continuum model 
(see parameter values in Table~\ref{tab:WA}).
}
\begin{tabular}{l|c|ccc|cccc}
\hline
\hline
\noalign {\smallskip}  
Source        &   N$_{\rm H}^{\rm Gal}$& $\Gamma_{2-5}$ & $\chi^{2}$/d.o.f  & P($\chi^{2}$)&  N$^{\rm in}_{\rm H}$&  $\Gamma_{0.3-2}$    & $\chi^{2}$/d.o.f&  P($\chi^{2}$) \\
\noalign {\smallskip}  
\hline
\noalign {\smallskip} 
I\,Zw1& 4.94 & 2.30$\pm$0.05 & 167.4/132 & 2.0 &$^{(a)}$7.8$\pm$1.0 & 2.63$\pm$0.06 & 352.8/324& 13.0\\ 
PG\,0804+761 & 3.01   &2.19$\pm$0.09   &161.5/196 & 96.6&  $^{(a)}$ 2.3$\pm$1.1  & 2.51$\pm$0.08 & 248.6/220   & 9.0 \\  
PG\,0947+396  & 1.55 &1.81$\pm$0.12 & 128.3/122 & 33.1  &$^{(a)}$ 2.1$\pm$1.4& 2.52$\pm$0.09&  285.1/312 & 86.0\\ 
PG\,0953+414  & 0.82 & 2.04$\pm$0.11 & 55.7/75 & 95.2 & $<$0.4 & 2.59$\pm$0.03 & 337.0/278 & 0.9 \\ 
PG\,1048+342  &  1.94   & 1.83$\pm$0.12 & 85.8/81 & 33.7  & $<$0.8 & 2.40$^{+0.06}_{-0.02}$  & 239.3/267 & 8.5 \\
PG\,1114+445 & 1.62 &  1.38$\pm$0.08 &  123.3/131& 67.2 & $^{(a)}<$2.0  & 1.56$^{+0.12}_{-0.03}$  &  267.9/213   & 0.6 \\
PG\,1115+407 &  1.91  &2.27$\pm$0.15 & 58.9/81 &96.9&$^{(a)}$ 2.1$^{+1.7}_{-1.3}$ & 2.92$\pm$0.09 & 268.0/300  & 90.8 \\ 
PG\,1116+215  &  1.29 & 2.26$\pm$0.16 & 98.6/80  & 7.8  & $<$0.5  & 2.71$\pm$0.04 & 263.6/294 &89.8   \\
PG\,1202+281 & 1.70 & 1.65$\pm$0.11 & 68.9/78 & 76.0 & $^{(a)}$ $<$0.3 & 2.28$\pm$0.03 & 309.9/264 & 2.7\\
PG\,1244+026  &   1.78  & 2.52$\pm$0.18 & 96.5/79 & 8.8&11.5$\pm$0.1  & 3.72$\pm$0.09  & 407.6/296 & 0 \\
PG\,1307+085  &  2.10   &  1.40$\pm$0.16 & 74.5/81 & 68.3 & $^{(a)}$ $<$0.3 & 2.22$\pm$0.05 & 268.8/219  & 1.2 \\
PG\,1309+355&1.02 & 1.76$\pm$0.18& 51.8/62&81.8&$^{(a)}$ $<$0.5 & 2.26$\pm$0.05 & 197.8/212  & 75.0 \\
PG\,1322+659  &  2.00  & 2.23$\pm$0.20 &52.8/57 &63.2  &  $<$0.7 &   2.99$^{+0.05}_{-0.03}$ &275.6/257 & 20.3    \\
PG\,1352+183  & 2.05   & 1.95$\pm$0.15 & 126.2/107 & 9.9 &   $^{(a)}$ $<$2.3  &  2.69$^{+0.10}_{-0.08}$& 311.9/299  & 29.2    \\
PG\,1402+261  & 1.47  &  2.31$\pm$0.16    & 75.5/73 &39.8  & $<$1.5 &2.83$^{+0.10}_{-0.03}$  &319.4/288 & 9.8  \\
Mrk\,1383 & 2.85 & 1.95$\pm$0.13 & 92.9/110 & 87.9 & $<$0.8 &2.62$^{+0.06}_{-0.02}$  & 321.2/313 & 36.2 \\      
PG\,1427+480  &  1.86  & 1.90$\pm$0.12 & 104.3/123 & 90.0 &  $<$1.4   &2.52$^{+0.08}_{-0.03}$ & 344.1/311   & 9.5  \\
Mrk\,478 & 1.04 & 2.45$\pm$0.13   & 90.9/115 & 95.2& 0.9$\pm$0.8  & 3.12$^{+0.07}_{-0.06}$& 322.7/319  & 43.2 \\ 
PG\,1512+370   &  1.35  & 1.72$\pm$0.13  & 125.0/124 &45.8& $<$0.8 & 2.31$\pm$0.04 & 301.4/305  & 54.7 \\
Mrk\,876     &  2.85  & 1.92$\pm$0.20 & 47.2/61 & 90.2 & $^{(a)}$ $<$2.3 & 2.45$^{+0.15}_{-0.04}$ & 254.2/230  & 13.1 \\
PG\,1626+554   &   2.02 &  2.03$\pm$0.21  &  61.2/82 & 95.8 &  $<$0.7 & 2.00$^{+0.19}_{-0.13}$ &  96.9/122  & 95.4    \\
\noalign {\smallskip} 
\hline
\hline
\end{tabular}
\label{table:fits}
\end{table*}
\begin{table*}[h!]
\caption{Spectral fit with a broken power law model ($\Gamma_{\rm soft}$, $\Gamma_{\rm hard}$, 
and E$_{\rm c}$ break point for the energy in keV), 
 including the Galactic (N$^{\rm Gal}_{\rm H}$) and intrinsic (N$^{\rm in}_{\rm H}$)
 absorptions (expressed in 10$^{20}$\,cm$^{-1}$).  
The unabsorbed fluxes are expressed in 10$^{-12}$\,erg\,s$^{-1}$\,cm$^{-2}$. 
 In parentheses are reported the luminosities (expressed in 10$^{44}$\,erg\,s$^{-1}$). 
 The last column reports the flux ratio in the 0.3--2\,keV and 0.3--10\,keV energy bands. 
 For PG\,1114+445, two strong absorption edges are required (see values in Table~\ref{tab:WA}).
}
\small 
\begin{tabular}{lccccccccccl}\hline\hline
\multicolumn{1}{c}{Source}   &      N$^{\rm in}_{\rm H}$   &     $\Gamma_{\rm soft}$        &   $\Gamma_{\rm hard}$     & E$_{\rm c}$    &$\chi^{2}$/d.o.f   & P($\chi^{2}$)   &  F$_{0.3-10}$ (L) &   F$_{2-10}$  (L) & $\frac{\rm 0.3-2\,keV}{0.3-10\,keV}$   \\
\noalign {\smallskip}  
\hline
\noalign {\smallskip}          
I\,Zw1  &  5.0$\pm$0.6  & 2.61$\pm$0.10 & 2.37$\pm$0.02 & 0.67$\pm$0.05 & 642.0/495 & 0 & 28.3 (2.1) & 8.4 (0.63) & 70$\%$\\      
PG\,0804+761& $<$ 2.8  & 2.47$^{+0.19}_{-0.10}$ & 2.16$^{+0.04}_{-0.09}$ &1.26$^{+0.50}_{-0.12}$& 489.0/475  & 31.8 & 32. (6.6)& 11. (2.2)& 66$\%$\\ 
PG\,0947+396  & $<$ 0.7  & 2.41$\pm$0.03 & 1.75$\pm$0.09  & 2.28$\pm$0.23  &  462.9/484&  74.8&   4.6  (4.3) & 1.8 (1.5) &  61$\%$  \\ 
PG\,0953+414  & $<$0.6 &  2.60$^{+0.05}_{-0.02}$  &  2.02$\pm$0.07  & 1.77$\pm$0.19 & 410.9/380  & 13.2 & 9.3 (12.3) & 3.0 (3.7) & 68$\%$  \\ 
PG\,1048+342  &  $<$1.2 & 2.40$^{+0.11}_{-0.03}$  & 1.80$\pm$0.07 & 1.80$^{+0.21}_{-0.17}$ & 428.3/386 & 6.8 & 3.3 (1.9) & 1.4 (0.76) & 58$\%$    \\
PG\,1114+445 &  $<$1.5 & 1.56$^{+0.09}_{-0.04}$ &  1.53$\pm$0.03 & 2.0 (f) &  481.4/410 &  0.9 &  3.1 (1.2) & 2.3  (0.89) &  26$\%$\\
PG\,1115+407 & $<$ 0.8 & 2.85$^{+0.06}_{-0.02}$ & 2.19$\pm$0.10   & 2.04$^{+0.24}_{-0.20}$ & 357.0/406 &  96.2 & 5.3 (2.9) & 1.2 (0.60) & 77$\%$ \\ 
PG\,1116+215  & $<$1.9  & 2.72$^{+0.18}_{-0.03}$& 2.17$\pm$0.11  & 1.76$^{+0.32}_{-0.41}$ &396.3/401   & 55.7  & 12. (8.3) & 3.2 (2.1) & 73$\%$      \\
PG\,1202+281 &  $<$0.5 & 2.29$\pm$0.04 & 1.71$\pm$0.06 & 1.76$^{+0.24}_{-0.16}$ & 413.8/375 & 8.2 & 7.8 (4.4)& 3.6 (1.9) & 55$\%$ \\
PG\,1244+026  & 12.1$\pm$0.1 &3.77$\pm$0.11  & 2.51$\pm$0.10 & 1.81$^{+0.12}_{-0.09}$   & 523.8/403 & 0   &  29. (1.4) &2.2 (0.099)   &  92$\%$\\ 
PG\,1307+085    & $<$7.0 &  2.86$^{+0.47}_{-0.22}$  & 1.52$\pm$0.06  & 0.93$\pm$0.06 & 360.4/338  & 19.3 & 3.4 (1.7) & 1.9 (0.87)    & 44$\%$  \\
PG\,1309+355 & $<$3.4 & 2.92$^{+0.34}_{-0.13}$ &1.73$\pm$0.05 & 0.78$\pm$0.05 & 269.1/301  & 90.7  & 1.4 (1.0)  & 0.71 (0.48)  & 49$\%$   \\
PG\,1322+659  & $<$2.8 & 3.01$^{+0.24}_{-0.04}$ & 2.18$^{+0.14}_{+0.11}$ &1.62$^{+0.19}_{-0.25}$ &337.3/336& 47.0 & 5.8 (3.8)  & 1.3 (0.77)  & 78$\%$ \\
PG\,1352+183  & $<$0.7   & 2.65$\pm$0.04    & 1.93$\pm$0.10    & 2.00$^{+0.17}_{-0.23}$& 485.2/458 & 18.3 & 6.0 (3.0)  & 1.9 (0.88)  & 68$\%$      \\
PG\,1402+261  & $<$ 2.8 & 2.91$^{+0.13}_{-0.09}$  & 2.18$^{+0.08}_{-0.13}$  & 1.73$^{+0.35}_{-0.17}$  & 424.1/384 & 7.7 &     7.7 (4.7)  & 1.8 (1.0) & 77$\%$    \\
Mrk\,1383  & $<$1.2 & 2.62$^{+0.11}_{-0.02}$ & 1.96$\pm$0.10 & 1.91$^{+0.25}_{-0.28}$ & 457.4/460 & 52.6 &23. (3.6) & 7.5 (1.1) & 67$\%$   \\
PG\,1427+480  & $<$1.6 &  2.53$^{+0.10}_{-0.03}$ & 1.87$\pm$0.07 & 1.92$^{+0.17}_{-0.22}$ &  495.5/488 &  39.8  &  2.8  (3.0)  & 1.0 (1.0) & 64$\%$\\
Mrk\,478 & 2.2$\pm$1.1&3.25$\pm$0.10 & 2.38$\pm$0.06 &1.56$\pm$0.10 & 422.3/465 & 92.2 & 17. (2.2) &2.7 (0.33) &84$\%$\\  
PG\,1512+370  &$<$ 0.9    & 2.31$\pm$0.04  & 1.78$\pm$0.05 &1.81$^{+0.16}_{-0.13}$   & 473.1/479 & 56.7     & 4.3 (13.)  &   1.9 (5.3) & 56$\%$\\
Mrk\,876    & $<$2.3     & 2.56$^{+0.19}_{-0.06}$  &  1.85$^{+0.08}_{-0.10}$ & 1.45$^{+0.21}_{-0.13}$  & 344.6/322  & 18.5  & 12.9 (4.5) & 5.2 (1.7) & 60$\%$ \\
PG\,1626+554 & $<$0.8  & 2.41$\pm$0.05& 1.99$\pm$0.12 & 1.72$^{+0.43}_{-0.52}$ & 388.0/388 & 49.0   & 8.3 (3.0)  &  3.1 (1.1) & 63$\%$ \\
\noalign {\smallskip}                       
\hline
\hline
\end{tabular}
\label{table:fitbkn}
\end{table*}

\clearpage
\begin{table*}[h!]
\caption{Spectral fit of absorption edge(s). $\tau$ is the optical depth. 
Significant detections ($\geq$ 99$\%$)  are marked in bold face. 
}
\begin{tabular}{l|ccc|ccc}
\hline
\hline
 \multicolumn{1}{c}{Object}  & $\tau_{\rm O VII}$  & $\tau_{\rm O VIII}$ & detection & \multicolumn{2}{c}{Edge parameters} & detection \\
\noalign {\smallskip}  
                             & (E=0.74\,keV) &  (E=0.87\,keV)&         &   E (keV)          &  $\tau$    & \\
\noalign {\smallskip}  
\hline
\noalign {\smallskip}  
I\,Zw1 & 0.11$\pm$0.04 & $<$0.01 & {\bf 99.9$\%$} & 0.65$\pm$0.02 & 0.21$\pm$0.04  & {\bf $>$99.99$\%$} \\
PG\,0804+761 &$<$0.09 & 0.07$\pm$0.06 & 90.3$\%$ & 0.62$\pm$0.04  &0.16$\pm$0.09 & {\bf 99.2$\%$}\\
             &       &               &         & 0.95$\pm$0.05  & 0.13$\pm$0.06\\
PG\,0947+396 & $<$0.10  & 0.15$\pm$0.08 & {\bf 99.8$\%$} & 0.85$\pm$0.04   & 0.16$\pm$0.08  & {\bf 99.9$\%$}\\
PG\,1114+445 & 2.26$^{+0.22}_{-0.19}$  &  0.32$\pm$0.16  & {\bf $>$99.99$\%$} & 0.72$\pm$0.01  & 2.23$\pm$0.17  & {\bf $>$99.99$\%$} \\
             &                    &                 &                   & 0.93$\pm$0.03  &  0.53$\pm$0.13 & \\
PG\,1115+407 & $<$0.19  & 0.08$\pm$0.06 & 91.8$\%$ & 0.78$\pm$0.05  & 0.15$\pm$0.08  & {\bf 99.4$\%$}\\
             &          &              &         & 1.18$\pm$0.10 &  0.15$\pm$0.07 & \\  
PG\,1116+215 & $<$0.19    & 0.09$\pm$0.07   & 97.7$\%$   & \\
PG\,1202+281 &  0.09$\pm$0.07  & $<$0.13 & {\bf 99.9$\%$}& 0.80$\pm$0.05   & 0.13$\pm$0.07 &  {\bf 99.92$\%$}  \\
 PG\,1307+085 & 0.33$^{+0.20}_{-0.16}$ & $<$0.44 & {\bf 99.97$\%$}&  0.78$\pm$0.04 & 0.39$\pm$0.16 &{\bf 99.97$\%$} \\
 PG\,1309+355&0.46$^{+0.20}_{-0.15}$ & $<$0.25 & {\bf 99.98$\%$} & 0.75$\pm$0.03 & 0.49$\pm$0.17  & {\bf 99.97$\%$}\\
PG\,1352+183&  0.11$\pm$0.08 & $<$0.08 & 97.0$\%$ & 0.72$\pm$0.06 & 0.12$\pm$0.08  & {\bf 99.7$\%$}\\
Mrk\,876 & $<$0.23  & 0.25$\pm$0.12 & {\bf 99.92$\%$} & 0.93$\pm$0.05 & 0.26$\pm$0.11   &  {\bf 99.97$\%$}\\
\noalign {\smallskip}  
\hline
\hline
\end{tabular}
\label{tab:WA}
\end{table*}
\clearpage 
\begin{table*}[h!]
\caption{Best-fitting spectral parameters for objects presenting a 
significant line detection (i.e. P$_{\rm prob}$ $\geq$ 90$\%$).
GA: Gaussian profile. {\sc D (diskline)} and {\sc L (laor)}: profile line
emitted by a relativistic accretion disk for a non-rotating BH  (Fabian et al. \cite{Fabian89}) and 
 a maximally rotating BH (Laor \cite{La91}), respectively. 
 The inclination of the disk is fixed to 30 deg.  
 We assume $q$= $-$2 (emissivity law),  (*) for PG\,1402+261 
 $q$ is let as a free parameter: $|q|>$3.0, 
and $|q|$=3.3$^{+0.3}_{-0.5}$, respectively for {\sc D} and {\sc L}.
 {\sc D}: R$_{\rm in}$=6\,R$_{\rm g}$ and R$_{\rm out}$=1\,000\,R$_{\rm g}$.
 {\sc L}: R$_{\rm in}$=1.26\,R$_{\rm g}$, R$_{\rm out}$=400\,R$_{\rm g}$. 
}
\vspace*{-0.2cm}
\begin{center}
\begin{tabular}{lllrrrr}
\hline
\hline
\noalign {\smallskip}                       
object    & line & \multicolumn{3}{c}{Line parameters} & \multicolumn{1}{c}{P$^{\rm line}_{\rm prob}$}& \multicolumn{1}{c}{P$^{\rm broad}_{\rm prob}$} \\ 
          &        & \multicolumn{1}{c}{E} & \multicolumn{1}{c}{$\sigma$} & \multicolumn{1}{c}{EW}  &\multicolumn{1}{c}{($\%$) }  &\multicolumn{1}{c}{($\%$) } \\ 
          &        &\multicolumn{1}{c}{(keV)} & \multicolumn{1}{c}{(keV)} & \multicolumn{1}{c}{(eV)}  &   & \\ 
\noalign {\smallskip}                       
\hline
\noalign {\smallskip}                       
 I\,Zw1 &   GA &   $~$6.89$^{+0.10}_{-0.15}$ & 0.01 (f)   &  69$\pm$35 &   99.2\\ 
      &  GA  & $~$6.58$^{+0.25}_{-0.37}$ & 0.60$^{+0.52}_{-0.30}$ &  297$^{+240}_{-141}$ & $>$99.9   & 99.8 \\
      &  {\sc D} & $~$6.78$\pm$0.13 &   & 180$\pm$68  & $>$99.9   \\
      &  {\sc L} & $~$6.74$^\pm$0.11 &   & 240$\pm$86  &  $>$99.9  \\
\noalign {\smallskip}                       
\hline
\noalign {\medskip}  
PG\,0804+761  &  GA &  $~$6.76$\pm$0.06 & 0.01 (f) & 140$^{+78}_{-76}$  & 98.5\\ 
       &  GA & $~$6.67$^{+0.31}_{-0.37}$ & 0.56$^{+0.55}_{-0.22}$ & 482$^{+356}_{-243}$&99.7 & 98.4 \\
              & {\sc D} & $~$6.82$^{+0.16}_{-0.27}$ &  &  320$^{+151}_{-156}$ & 99.6 \\
              & {\sc L} & $~$6.71$^{+0.27}_{-0.17}$ &  & 400$^{+174}_{-180}$ & 99.8\\
\noalign {\smallskip}                       
\hline
\noalign {\smallskip}                       
 PG\,0947+396 &   GA &  $~$6.45$^{+0.07}_{-0.12}$ & 0.01 (f)    & 92$\pm$65 & 94.9 \\ 
\noalign {\smallskip}                       
\hline
\noalign {\smallskip}                       
 PG\,1048+342 &   GA &  $~$6.35$^{+0.07}_{-0.10}$ & 0.01 (f) & 123$\pm$64 & 98.3\\ 
\noalign {\smallskip}                       
\hline
\noalign {\smallskip}                       
 PG\,1114+445  & GA  & $~$6.40$^{+0.02}_{-0.05}$  & 0.01 (f)  & 125$\pm$41    & $>$99.9\\  
\noalign {\smallskip}                       
\hline
\noalign {\smallskip}                       
 PG\,1115+407 &   GA & $~$7.20$^{+0.13}_{-0.45}$ & 0.01 (f)        & 157$^{+115}_{-125}$ & 92.4 \\ 
\noalign {\smallskip}                       
\hline
\noalign {\smallskip}                       
PG\,1116+215  &  GA      & $~$7.23$^{+0.07}_{-0.30}$ & 0.01 (f) & 194$\pm$123  & 96.3 \\ 
            &  GA  & $~$6.67$^{+0.36}_{-0.41}$ & 0.72$^{+0.48}_{-0.31}$ & 885$^{+563}_{-213}$ & 99.8 & 99.7\\
               &  {\sc D} & $~$7.10$^{+0.14}_{-0.16}$ &  & 517$^{+251}_{-237}$ &  99.8\\
               &  {\sc L} & $~$7.06$\pm$0.18 &  & 636$^{+306}_{-285}$ &  99.9\\
\noalign {\smallskip}                       
\hline
\noalign {\smallskip}                       
 PG\,1244+026 &  GA      & $~$6.66$^{+0.09}_{-0.07}$   & 0.01 (f)   & 222$\pm$156   & 92.4 \\
\noalign {\smallskip}                       
\hline
\noalign {\smallskip}                       
 PG\,1309+355 &   GA &  $~$6.40$^{+0.10}_{-0.12}$ & 0.01 (f)  & 156$^{+100}_{-104}$ & 97.4\\ 
      &  GA  & $~$6.48$^{+0.18}_{-0.23}$ & 0.21$^{+0.24}_{-0.10}$ & 300$^{+206}_{-173}$ & 98.7  & 94.1  \\
             &  {\sc D}  & $~$6.54$^{+0.10}_{-0.16}$ &  & 373$^{+187}_{-192}$ &  99.8 \\
            &  {\sc L}  & $~$6.51$^{+0.11}_{-0.15}$ &  & 458$\pm$237 & 99.7 \\
\noalign {\smallskip}                       
\hline
\noalign {\smallskip}                       
 PG\,1402+261 &   GA & $~$7.38$^{+0.07}_{-0.10}$ & 0.01 (f) & 262$^{+139}_{-140}$ &  98.5\\ 
          &GA&$~$7.26$^{+0.44}_{-0.51}$ &1.18$^{+0.63}_{-0.40}$&$~$2040$^{+1340}_{-880}$& $>$99.9 & $>$99.9\\ 
             &  {\sc D}$^{(*)}$  & $~$8.65$^{+0.53}_{-0.37}$ &  & 2000$^{+700}_{-780}$& $>$99.9 \\
            &  {\sc L}$^{(*)}$  & $~$8.51$^{+0.33}_{-0.24}$ &  & 4460$^{+2050}_{-1820}$  &  $>$99.9 \\
\noalign {\smallskip}                       
\hline
\noalign {\smallskip}    
Mrk\,1383     & GA    & $~$6.62$^{+0.12}_{-0.10}$   & 0.01 (f)  & 107$^{+72}_{-81}$ &   91.5\\      
\noalign {\smallskip}    
\hline
\noalign {\smallskip}                       
 PG\,1512+370 &   GA &  $~$6.51$\pm$0.13 & 0.01 (f)                  & 93$\pm$64 & 94.4 \\ 
\noalign {\smallskip}                       
\hline
\hline
\end{tabular}
\end{center}
\label{tab:FeK}
\end{table*}
\clearpage

\begin{table*}[h!]
\caption{Fit results, in the 2--12\,keV energy band, of a 
 disk reflection model assuming a {\sc lamppost} geometry for I\,Zw1, 
PG\,0804+761, PG\,1114+445, PG\,1116+215 and PG\,1309+355, 
i.e. we assume a height of 10\,R$_{g}$, a ratio of X-ray to disk flux of 0.2 
 (with solar iron abundance).
For PG\,1402+261, a lower height above the AD of about 2.5\,R$_{g}$, 
an iron abundance of 5 relative to solar abundance are required. 
Its spectrum is dominated by reflection. 
}
\begin{center}
\begin{tabular}{ccccccc}
\hline
\hline
\noalign {\smallskip}  
Object &  $\Gamma$  & $\theta$ (deg) & r$_{\rm in}$ (r$_{\rm g}$) & $\chi^{2}$/dof \\  
\noalign {\smallskip}  
\hline
\noalign {\smallskip}  
 I\,Zw1      &  2.28$\pm$0.03    & 42$^{+23}_{-19}$ & 16$^{+17}_{-10}$& 598.7/590 \\
\noalign {\smallskip}  
PG\,0804+761 & 2.07$\pm$0.06     & 42$\pm$21  &$<$14  & 212.2/250\\
\noalign {\smallskip}  
PG\,1116+215 & 2.12$\pm$0.09 & 30$\pm$22 & $<$13 & 124.6/104 \\ 
\noalign {\smallskip}  
PG\,1309+355 & 1.74$\pm$0.09 & $<$27 &$<$30 & 62.9/86 \\
\noalign {\smallskip}  
PG\,1402+261 & 2.13$\pm$0.10 & $>$60 & $<$5 & 85.9/92\\
\noalign {\smallskip}  
\hline
\hline
\end{tabular}
\end{center}
\label{tab:lamppost}
\end{table*}

\begin{table*}[h!]
\caption{Spectral fit with  an absorbed black body spectrum 
of an accretion disk ({\sc diskpn}, kT$_{\rm diskpn}$) combined with 
a power law model.
T$^{\rm max}_{\rm disk}$ (in eV) represents the maximal values 
of the disk temperature for a standard $\alpha$ thin accretion disk at 3\,R$_{\rm S}$. 
 kT$_{\rm diskpn}$ is expressed in eV. 
 (a): one or two absorption edges are  added to the continuum model due to the presence of a WA.
(b): a Fe\,K line is added (either a Gaussian line or a {\sc Laor} line).
}
\begin{tabular}{l@{}ccccc}
\hline
\noalign {\smallskip}  
\multicolumn{1}{c}{Source}   & T$^{\rm max}_{\rm disk}$ & kT$_{\rm diskpn}$  &  $\Gamma$  & $\chi^{2}$/d.o.f \\
\noalign {\smallskip}  
\hline
\noalign {\smallskip}               
 I\,Zw1$^{(a,b)}$    & 39  & 200$\pm$25  & 2.33$\pm$0.03    & 540.0/493 \\
PG\,0804+761$^{(a,b)}$& 15 &  159$^{+30}_{-23}$ & 2.22$\pm$0.07 & 464.1/468  \\
PG\,0947+396$^{(a,b)}$  & 11 & 241$\pm$15 & 1.88$\pm$0.06 & 451.0/480 \\  
PG\,0953+414  & 18 &  174$\pm$8 & 2.12$\pm$0.04 & 421.9/380   \\ 
PG\,1048+342$^{(b)}$  & 15 & 183$\pm$11 & 1.91$\pm$0.05 & 423.2/384\\
PG\,1115+407$^{(a,b)}$ & 24   & 172$\pm$14 & 2.40$\pm$0.07 & 353.1/402   \\
PG\,1116+215$^{(a,b)}$  & 13 & 175$^{+17}_{-26}$  & 2.30$\pm$0.07 & 375.8/397  \\             
PG\,1202+281$^{(a)}$  & 14 & 223$^{+14}_{-17}$ &  1.72$\pm$0.05  & 396.4/373   \\
PG\,1244+026$^{(b)}$  & 73 & 197$\pm$11  & 2.39$\pm$0.10   & 427.5/401  \\
PG\,1307+085$^{(a)}$  & 9   & 142$^{+41}_{-19}$ & 1.55$\pm$0.07 & 338.6/336 \\
PG\,1309+355$^{(a,b)}$  &16   & 144$^{+39}_{-26}$ & 1.78$\pm$0.08  & 244.4/296 \\
PG\,1322+659          &19 &     149$\pm$8 & 2.28$\pm$0.07   & 226.3/336    \\
PG\,1352+183$^{(a)}$  &  15    & 201$\pm$12 & 2.04$\pm$0.08  & 478.8/456 \\
PG\,1402+261$^{(b)}$  & 31  & 151$^{+11}_{-17}$ & 2.53$\pm$0.09 & 400.4/381 \\ 
Mrk\,1383$^{(b)}$   & 7 & 160$\pm$8  & 2.09$\pm$0.05   & 442.8/458 \\
PG\,1427+480  & 21&     190$\pm$9  & 1.98$\pm$0.05 & 505.9/488   \\
 Mrk\,478   & 35 &  138$\pm$4   &  2.48$\pm$0.05  & 455.0/465   \\
PG\,1512+370$^{(b)}$  & 9  &    210$\pm$19 & 1.90$\pm$0.05 & 490.7/477 \\
Mrk\,876$^{(a)}$    & 9   & 177$\pm$24  & 1.87$\pm$0.11 & 326.9/320 \\
PG\,1626+554   & 12 & 145$^{+22}_{-24}$  & 2.09$\pm$0.07     & 388.5/388   \\
\noalign {\smallskip}                       
\hline
\hline
\end{tabular}
\label{table:fitbb}
\end{table*}

\clearpage
\begin{table*}[h!]
\caption{Spectral fit with absorbed compTT+power law. For the {\sc compTT} component model, 
the soft photon temperature (kT$^{\rm photon}_{1}$) is fixed to  Tmax (column 1 of Table~\ref{table:fitbb}). 
 kT$^{\rm plasma}$ is expressed in keV. 
$\tau$ corresponds to the plasma optical depth. 
The fits include the Galactic (N$^{\rm Gal}_{\rm H}$) and intrinsic (N$^{\rm in}_{\rm H}$) absorptions. 
Column densities are expressed in 10$^{20}$\,cm$^{-1}$. 
The unabsorbed fluxes are expressed in 10$^{-12}$\,erg\,s$^{-1}$\,cm$^{-2}$
 In parentheses are reported the luminosities (expressed in 10$^{44}$\,erg\,s$^{-1}$). 
 The last column reports the flux ratio in the 0.3--2\,keV and 0.3--10\,keV energy bands. 
 (a): one or two absorption edges are  added to the continuum model due to the presence of a WA.
(b): a Fe\,K line is added (either a Gaussian line or a {\sc Laor} line).
 }
\begin{tabular}{lcccccccccc}\hline
\multicolumn{1}{c}{Source} &    N$^{\rm in}_{\rm H}$  &   kT$^{\rm plasma}$    &  $\tau_{1}$    &  $\Gamma$   &   $\chi^{2}$/d.o.f   &   F$_{0.3-10}$ (L)   &   F$_{2-10}$ (L) & $\frac{\rm 0.3-2\,keV}{0.3-10\,keV}$ \\
\noalign {\smallskip}  
\hline
\noalign {\smallskip}            
  I\,Zw1$^{(a,b)}$ &  10.5$\pm$2.6 & 0.22$\pm$0.05 & 14.7$^{+2.2}_{-1.3}$ & 2.33$\pm$0.03 & 525.7/490 & 36.3 (2.8) & 8.6 (0.64) & 76$\%$\\
PG\,0804+761$^{(a,b)}$&  3.9$^{+7.2}_{-3.1}$& 2.8$^{+3.1}_{-1.4}$& 3.5$^{+2.1}_{-1.1}$&1.53$^{+0.32}_{-0.17}$& 462.5/470 & 30.5 (6.1) & 11.4 (2.3) &63$\%$\\
PG\,0947+396$^{(a,b)}$& $<$3.3 & 0.32$\pm$0.06& 15.6$^{+0.8}_{-2.0}$ & 1.76$\pm$0.07  &  440.1/479   & 4.6 (4.2)  & 1.8 (1.6) &61$\%$ \\ 
PG\,0953+414  & $<$2.0 &  0.29$^{+0.10}_{-0.06}$ & 13.8$^{+0.5}_{-1.0}$ & 1.98$\pm$0.05  & 408.6/379  & 9.1 (11.9) & 3.0 (3.7) & 67$\%$  \\ 
PG\,1048+342$^{(b)}$  &  $<$ 1.8 & 0.25$^{+0.09}_{-0.06}$ &  16.6$^{+0.8}_{-1.5}$& 1.82$\pm$0.06  & 417.4/383  & 3.2 (1.9) & 1.3 (0.74) & 60$\%$  \\
PG\,1114+445$^{(a,b)}$ &5.2$^{+2.5}_{-3.5}$ & 3.1$^{+9.7}_{-2.2}$   &  6.7$^{+3.2}_{-4.0}$ & 1.51 (f)  & 214.6/236 & 3.3 (1.3) & 2.3 (0.92) & 30$\%$\\
PG\,1115+407$^{(a,b)}$ & $<$7.2 &  0.17$^{+0.08}_{-0.04}$  & 19.2$^{+5.4}_{-3.4}$ & 2.40$\pm$0.06 &333.2/399 & 5.8 (3.1) & 1.2 (0.60) & 79$\%$ \\ 
PG\,1116+215$^{(a,b)}$ & $<$11.8 & 0.25$^{+0.18}_{-0.07}$&13.1$^{+3.7}_{-2.1}$ & 2.23$\pm$0.08 & 370.3/396 & 14.8 (11.0)&3.2 (2.1) & 78$\%$ \\
PG\,1202+281$^{(a)}$ &$<$6.7 & 0.24$\pm$0.07 & 21.2$^{+1.8}_{-4.8}$& 1.68$\pm$0.04 &396.9/372 & 7.5 (4.2) & 3.6 (1.9) &52$\%$\\ 
PG\,1244+026$^{(b)}$ & $<$2.1 &0.18$\pm$0.02 & 28.7$^{+2.5}_{-4.0}$&  2.40$\pm$0.08 &  426.8/400 & 12.4 (0.57) &  2.1 (0.098)  & 83$\%$ \\
PG\,1307+085$^{(a)}$ & 5.4$\pm$4.2& 0.10$\pm$0.02 & $>$44 &  1.46$\pm$0.03 &  350.1/337  & 3.4 (1.6) & 1.9 (0.87) & 44$\%$  \\
PG\,1309+355$^{(a,b)}$& $<$13.1 & 0.13$^{+0.08}_{-0.03}$  & $>$23.4 & 1.78$^{+0.09}_{-0.06}$ & 244.4/295 & 1.4 (1.0) & 0.72 (0.48)  & 49$\%$   \\
PG\,1322+659 & $<$4.3 & 0.17$\pm$0.05 & 22.4$^{+1.7}_{-3.7}$  & 2.21$\pm$0.07 & 327.1/335  & 5.3 (3.4) & 1.3 (0.77) & 76$\%$      \\
PG\,1352+183$^{(a)}$ & $<$6.8  & 0.25$\pm$0.06   & 17.6$^{+0.9}_{-3.0}$ & 1.90$\pm$0.08 & 476.0/455 & 5.7 (2.8) & 1.9 (0.88) & 67$\%$\\
PG\,1402+261$^{(b)}$  & $<$3.1 & 0.25$^{+0.11}_{-0.06}$ & 14.2$^{+0.6}_{-1.7}$ & 2.20$\pm$0.08 & 407.5/381 &  7.3 (4.4)   & 1.8 (1.0)  & 75$\%$  \\
Mrk\,1383$^{(b)}$ & $<$ 1.0 & 0.20$\pm$0.05 & 19.1$^{+0.9}_{-1.2}$  & 2.02$\pm$0.05  & 438.8/457 & 23.2 (3.5) & 7.5 (1.1) &  68$\%$  \\
PG\,1427+480 & $<$4.8  & 0.27$^{+0.08}_{-0.05}$ & 16.0$^{+0.7}_{-2.4}$ & 1.86$\pm$0.05 & 494.9/487 & 2.7  (2.8)   & 1.0 (1.0)  &  63$\%$ \\ 
Mrk\,478  &  $<$5.1 &  0.22$\pm$0.05 & 13.6$\pm$1.3  & 2.29$\pm$0.06  & 416.4/464 & 15.8 (2.0) & 2.7 (0.33) & 83$\%$\\
PG\,1512+370$^{(b)}$& $<$6.5 &  0.44$^{+0.72}_{-0.12}$ & 10.9$^{+0.6}_{-2.4}$& 1.74$\pm$0.08 & 468.6/476 & 4.3 (13.5) &1.9 (5.4) &56$\%$ \\
Mrk\,876$^{(a)}$   &$<$4.0 & 0.14$\pm$0.06  & $>$ 37.4  & 1.86$\pm$0.06  & 327.4/320  & 11.8 (4.0)  & 5.1 (1.7)  & 57$\%$  \\
PG\,1626+554 &  $<$6.9 &   0.52$^{+7.84}_{-0.30}$  & 9.4$^{+0.7}_{-2.2}$ & 1.85$\pm$0.17   &  385.9/387 & 8.0 (2.9)& 3.1 (1.1) & 61$\%$\\
\noalign {\smallskip}                       
\hline
\hline
\end{tabular}
\label{table:fitcomptt}
\end{table*}

\clearpage
\begin{table*}[h!]
\caption{Mean and standard deviation ($\sigma$) 
of the photon index in the 0.3--2\,keV and 2--5\,keV energy ranges.
NLG: Narrow Line Galaxies (FWHM$<$2000\,km.s$^{-1}$), RQQ: radio-quiet quasars, and
RLQ (radio-loud quasars).}
\begin{center}
\begin{tabular}{ccc}
\noalign {\smallskip}  
 \hline
\hline
 Type                           &        0.3--2\,keV              & 2--5\,keV\\
\noalign {\smallskip}  
 \hline
\noalign {\smallskip}  
All objects                   &    2.56 ($\sigma$:0.44)    &  1.99 ($\sigma$:0.31) \\
\noalign {\smallskip}  
NLG                             & 3.04 ($\sigma$:0.42)          &  2.37  ($\sigma$:0.11) \\
\noalign {\smallskip}  
RQQ                            & 2.43 ($\sigma$:0.35)          &  1.90  ($\sigma$:0.27) \\
\noalign {\smallskip}  
RLQ                            & 2.29 ($\sigma$:0.04)           & 1.74 ($\sigma$:0.03)   \\
\hline
\hline
\end{tabular}
\end{center}
\label{tab:mean}
\end{table*}


\begin{table*}[h!]
\caption{Spearman-rank correlations. (r$_{\rm s}$, P$_{\rm rs}$): 
r$_{\rm s}$ Spearkman coefficient, and P$_{\rm rs}$ probability.
r$_{\rm s}>$0 and  r$_{\rm s}<$0 means a correlation and an anti-correlation, respectively. 
Significant correlations ($\geq$ 99$\%$)  are marked in bold face. }
\small
\begin{tabular}{ccccccc}
\hline
\hline
                                    &   $\Gamma_{\rm 2-5}$& L(2-10)           & $\Gamma_{\rm 0.3-2}$& L(0.3-2)     &   z  &FWHM \\
\noalign {\smallskip}  
 L(2-10)                        &        (-0.31,83.6)          &    -                    &           -                          &        -       &    -      &    -\\
\noalign {\smallskip}  
$\Gamma_{\rm 0.3-2}$& {\bf (0.85,$>$99.99)}    & (-0.35,87.8)    &           -                           &      -        &    - &    - \\
\noalign {\smallskip}  
L(0.3-2)                          &   (0.10,33.8)               &   {\bf  (0.86,$>$99.99)}   &    (0.12,39.5)    &       -       &  -  &  -  \\
\noalign {\smallskip}  
z                                     &   (-0.45,95.8)           &  (0.42,94.0)     &  (-0.24,71.0)                  &    (0.39,91.6)   & -& -\\
\noalign {\smallskip}  
FWHM                            & {\bf  (-0.66,99.9)}            & {\bf  (0.66,99.9)}  &   {\bf (-0.59,99.5)}       & (0.39,91.6)       &   (0.25,72.0) & -\\
\noalign {\smallskip}  
 R$_{\rm L}$                  &   (0.21,63.2)        &   (0.29,79.6)        &  (0.15,48.4)                       &  (0.38,91.2)       &  (0.03,10.5) & (0.07,24.2)\\
\noalign {\smallskip}  
M$_{\rm V}$                 &   (0.28,77.4)            &   (-0.54,98.9)       &  (0.19,58.5)       &   (-0.53,98.7)    &  (-0.36,89.0)  & (-0.24,71.0)\\
\noalign {\smallskip}  
M$_{\rm BH}$              &  {\bf (-0.65,99.8)}      &  {\bf (0.61,99.7)}      &   {\bf (-0.57,99.4)}       &  (0.30,81.8)     &(0.29,79.1)  & {\bf (0.72,99.98)}\\
\noalign {\smallskip}  
$\dot{m}$            &   {\bf (0.75,$>$99.99)}    &  (-0.41,93.4)        &   {\bf (0.71,99.97)}        &  (-0.05,17.7)   &(-0.09,29.0)   & {\bf (-0.76,$>$99.99)}\\
\noalign {\smallskip}  
\hline
\hline
\end{tabular}
\label{tab:corr}
\end{table*}
\clearpage
\appendix
\section{Notes on individual objects}\label{sec:notes}

\begin{table}[p!]
\caption{Summary for all objects for detections of soft excess (SE), warm absorber (WA), and Fe\,K${\alpha}$ line.
}
\begin{center}
\begin{tabular}{lccl}
\hline
\hline
\noalign{\smallskip}
\multicolumn{1}{c}{Source}     & SE                             &WA &   \multicolumn{1}{c}{Fe\,K$_{\alpha}$ line}  \\
\hline
\noalign{\smallskip}
\multicolumn{4}{c}{Narrow line galaxies  (5 objects)} \\
\noalign{\smallskip}
\hline
\noalign{\smallskip}
I\,Zw1         &                             & Yes                              & Yes (broad at $\sim$6.7\,keV) \\
\noalign{\smallskip}
PG\,1115+407       &Yes                         &                                & Yes (broad at $\sim$7\,keV) \\
\noalign{\smallskip}
PG\,1244+026        &Yes                         &                               &  Yes (narrow at $\sim$ 6.7\,keV)\\
\noalign{\smallskip}
PG\,1402+261        & Yes                          &                             &   Yes (blueshifted above 7\,keV)\\
\noalign{\smallskip}
Mrk\,478     & Yes                         &                                        &  \\
\noalign{\smallskip}
\hline
\hline
\multicolumn{4}{c}{Radio-quiet quasars   (14 objects)} \\
\noalign{\smallskip}
\hline
\noalign{\smallskip}
PG\,0804+761        &                          &       Yes                          & Yes (broad at $\sim$6.7\,keV) \\
\noalign{\smallskip}
PG\,0947+396        &Yes                      & Yes                               & Yes (narrow at 6.45\,keV) \\
\noalign{\smallskip}
PG\,0953+414        &Yes                         &                                      &  \\
\noalign{\smallskip}
PG\,1048+342       &Yes                         &                                      & Yes (narrow at 6.4\,keV)\\
\noalign{\smallskip}
PG\,1114+445       &                                  &     Yes         &         Yes (narrow at $\sim$ 6.4\,keV)    \\
\noalign{\smallskip}
PG\,1116+215       &Yes                         & Yes                               &   Yes (broad at $\sim$7\,keV) \\
\noalign{\smallskip}
PG\,1202+281        &Yes                         &  Yes                                &   \\
\noalign{\smallskip}
PG\,1307+085        &Yes                        & Yes                                 &   \\
\noalign{\smallskip}
PG\,1322+659       &Yes                        &                                          &    \\
\noalign{\smallskip}
PG\,1352+183        & Yes                             &                                          &    \\
\noalign{\smallskip}
Mrk\,1383  &Yes                         &                                    &  Yes (narrow at $\sim$ 6.6\,keV)  \\
\noalign{\smallskip}
PG\,1427+480      &Yes                         &                                             &  \\
\noalign{\smallskip}
Mrk\,876  & Yes                      &       Yes                                 & \\
\noalign{\smallskip}
PG\,1626+554       &   Yes                  &                                                    & \\
\noalign{\smallskip}
\hline
\hline
\multicolumn{4}{c}{Radio-loud quasars  (2 objects)} \\
\noalign{\smallskip}
\hline
PG\,1309+355        &Yes                        &  Yes                                   &  Yes (narrow at 6.4\,keV) \\
\noalign{\smallskip}
PG\,1512+370       &Yes                       &                                                 & Yes (narrow at 6.5\,keV) \\
\noalign{\smallskip}
\hline
\hline
\end{tabular}
\end{center}
\label{tab:sum}
\end{table}

In this section, we give some notes about each object in our sample. 
 An exhaustive comparison with previous observations for each objects 
is beyond the scope of the work. 
For an overview of their X-ray characteristics found with {\sl XMM-Newton}, 
  see Table~\ref{tab:sum} where we report
 for each object the possible  presence of a soft excess, 
of a warm absorber, and of a Fe\,K line. \\

\indent All objects of this sample were already observed by {\sl ROSAT} 
 (Wang et al. \cite{WBB96}). 
 The soft photon indices found during the {\sl XMM-Newton} observations 
(see Table~\ref{table:fits}) are consistent within the errors 
for most of the quasars with the {\sl ROSAT} values reported 
 by Wang et al. (\cite{WBB96}), except for PG\,1114+445, PG\,1244+026, and PG\,1307+085.

\subsection{I\,Zw1 (z=0.061)}

I\,Zw1 is the prototype "Narrow Line Seyfert 1" nucleus whose spectrum 
is widely used as an \ion{Fe}{ii} template for almost all quasars. 
This object has already been observed with {\sl ASCA}  (Reeves \& Turner \cite{RT2000})  
 and was well fitted with an absorbed power law model (over the 0.5--10\,keV energy range) 
combined with a narrow Gaussian emission line. 
 They found  a very soft photon index of $\Gamma$=2.37$\pm$0.05,  
 and a very large EW of 483$\pm$212\,eV at 6.77$^{+0.11}_{-0.17}$\,keV.
 In their data a broad line was only marginally required at a probability
 of 94.6$\%$. 
 In the present {\sl XMM-Newton} observation, we found that 
a broad  line is statistically required with a probability of
 99.8$\%$ compared to a narrow line. 
 This line has one of the highest significances of our sample with
 a probability higher than 99.9$\%$. 
 The energy of the line found here is consistent within the errors 
with the Reeves \& Turner (\cite{RT2000}) result.   
Recently, Gallo et al. (\cite{Ga2004})  presented a detailed {\sl XMM-Newton}
 analysis of I\,Zw1. As in the present work, they found signatures  
of a warm absorber. They confirmed the presence of a prominent  iron feature,
 and suggest that  the broad emission feature can be attributed to a neutral
 Fe K${\alpha}$ line in addition to a blend of He- and H-like Fe K${\alpha}$ 
lines.  The light curve shows a strong, hard  X-ray flare mainly in the 
3-12 keV band.  The timing results are consistent with the flare originating
 in the accretion disk corona. However the iron emission line(s) did not
 appear to respond to changes of the continuum flux during the flare. 

\subsection{PG\,0804+761 (z=0.100)}

PG\,0804+761 was one of the brightest radio-quiet quasars observed by {\sl ASCA}
 with a energy spectrum well represented by a simple power law model with 
a photon index of 2.18$^{+0.02}_{-0.03}$ in the 2--10\,keV energy range 
 with no significant Fe\,K emission detected (George et al. \cite{G2000}). 
Recently, Papadakis et al (\cite{Pap2003}) found similar results with
 {\sl RXTE} analyzing the spectrum averaged over one year 
($\Gamma_{\rm 3-15\,keV}$=2.10$^{+0.12}_{-0.10}$ and no  Fe\,K line is 
statistically required). The flux variations appeared not to be associated
 with any spectral variations.  During the present {\sl XMM-Newton} 
observation, we find in the 2--5\,keV band a consistent photon index of
 $\Gamma$=2.19$\pm$0.09. Combining  a narrow Gaussian emission line at
 6.4\,keV with an absorbed power law model, we find only weak evidence for
 such a line (F-test probability of only 92.8$\%$). 
 However, as reported in Table~\ref{tab:FeK}, leaving the energy of the line
 to vary, we find a much better fit for a narrow line at E=6.76$\pm$0.06\,keV
 (F-test probability of 98.5$\%$).   
 Leaving the width of the line to vary gives a probability of 99.7$\%$ for the 
detection of the line. The broad-band continuum is very well fitted by a
 broken power law model, the combination of a black body and a power law
 model, and best fitted by  a double Comptonization model. 

\subsection{PG\,0947+396 (z=0.206)}

PG\,0947+396 was observed previously by {\sl BeppoSAX} and its spectrum shown
 the presence of a line at 6.35$\pm$0.13\,keV and a large EW of
 670$\pm$170\,eV, however the reduced $\chi^{2}$=1.44 was unacceptable  
($\Gamma_{\rm 1.3-10\,keV}$=1.95$\pm$0.10, Table~6 in Mineo et al.
 \cite{M2000}). We find for the {\sl XMM-Newton} observation that adding a
 narrow line marginally improves the fit (F-test probability of 94.9$\%$). 
The energy of the line at 6.45$^{+0.07}_{-0.12}$\,keV is compatible with
 neutral to moderately ionized iron (i.e. \ion{Fe}{xvii}). 
The equivalent width of 92$\pm$65\,eV is compatible with those observed in
 RQQ (i.e. a mean EW of 163$\pm$17\,eV, Reeves \& Turner \cite{RT2000}). 
 The broad-band continuum is very well fitted either by a broken power law
 model or by  a double Comptonization model. We find no additional significant
 intrinsic absorption as expected from the weak UV absorption observed in this
 object (Brandt et al. \cite{B2000}). 

\subsection{PG\,0953+414 (z=0.239)}

PG\,0953+414 was observed by {\sl ASCA} (George et al. \cite{G2000}) in 
November 1997. They found that the spectrum was well fitted by an absorbed
 power law model with $\Gamma$=2.03$^{+0.06}_{-0.05}$. We find that the 
2--5\,keV spectrum is well fitted by an absorbed power law model with 
$\Gamma$=2.04$\pm$0.11, i.e. consistent with the previous observation. 
Additionally, we found the presence of a soft excess. The broad-band continuum
 is very well fitted either by a broken power law model, or by a double 
Comptonization model. 

\subsection{PG\,1048+342 (z=0.167)}

PG\,1048+342 was observed by {\sl BeppoSAX} (Mineo et al. \cite{M2000}) in May
 1997. They found no indication for the presence of a soft excess below 
2\,keV, and the spectrum was well fitted with an absorbed power law model
 ($\Gamma$=1.82$\pm$0.07). We find the presence of a soft excess and that the
 broad-band continuum is very well fitted either by a broken power law model, 
 or by  a double Comptonization model. 

\subsection{PG\,1114+445 (z=0.144)}

PG\,1114+445 was one of the few quasars for which the {\sl ROSAT} PSPC spectrum
could not be adequately represented by a single power law model 
($\Gamma$=2.36$\pm$0.49, Wang et al. \cite{WBB96}) and the only quasar
 (from 23 objects) for which there was strong evidence of absorption by
 ionized material (Laor et al. \cite{La94}). 
This X-ray absorption feature was also confirmed using  {\sl ASCA}  
 by George et al. (\cite{G97}) who found that the column density of
 the photoionized medium is about 2$\times$10$^{22}$\,cm$^{-2}$ with 
an ionization parameter of U$_{\rm x}\sim$0.1
  (U$_{\rm x}$ is defined in equation 1 in Netzer \cite{Ne96}).  
Such parameters are similar
 to those found in Seyfert\,1 galaxies at lower redshift. 
  UV absorption lines of Ly${\alpha}$, \ion{C}{iv}, and \ion{N}{v} 
were observed with  {\sl HST}  by Mathur et al. (\cite{Ma98}), 
who concluded that the mass outflow rate is comparable to the
 accretion rate in this object. \\
\indent The {\sl XMM-Newton} observation allows us to obtain a very good S/N 
spectrum of this object. We confirm the huge absorption in the soft X-rays
 which is satisfactorily fitted with two absorption edges of \ion{O}{vii}
 (0.74\,keV) and \ion{O}{viii} (0.87\,keV) with optical depths
of  about 2.3 and 0.3, respectively. However the determination of the exact
 nature of this medium (ionization state, velocity, etc...) needs data with
 higher spectral resolution such as from the RGS aboard {\sl XMM-Newton}. 
For exemple, the absorption feature is certainly a complex mixture of 
absorption lines and edges of Oxygen and Iron as the so-called UTA
 (Behar et al. \cite{Be2001}). Unfortunatly for the present exposure time the
 S/N of the RGS data is very poor with only about 700 counts 
(source + background) in each RGS instrument which is too low for any 
spectral analysis. Therefore, a longer observation of this object with 
spectrometers is required to probe the nature of the warm absorbing medium. 

\subsection{PG\,1115+407 (z=0.154)}

PG\,1115+407 is one of the four objects of our sample which displays a narrow 
 optical H$_{\beta}$ line at 5100\AA~ with a FWHM of 1720\,km\,s$^{-1}$, 
and therefore could be associated with a narrow line QSO similarly to NLS1
 galaxies. Mineo et al. (\cite{M2000}) found that the {\sl BeppoSAX} spectrum
 of this object displays a line at 6.69$\pm$0.11\,keV (EW=580$\pm$280\,eV) 
 similar to those found in NLS1 galaxies and consistent  with an origin from
 ionized gas. The {\sl XMM-Newton} data marginally require the presence of an
 iron K${\alpha}$ line near 7.2\,keV. This energy is compatible within the
 errors with the previous {\sl ASCA} observation, although at this energy the
 line may be due to higher ionized iron such as \ion{Fe}{xxvi} (H-like).

\subsection{PG\,1116+215 (z=0.177)}

 From an  {\sl ASCA} observation, Nandra et al. (\cite{N96}) found a high
 significance for a narrow line at E=6.90$\pm$0.15\,keV with a large EW of 
280$^{+160}_{-170}$\,eV. The centroid energy of this line corresponds to very
 highly ionized iron (\ion{Fe}{xxv}--\ion{Fe}{xxvi}) and could imply that the
 accretion disk is ionized as expected from the high ratio UV/X-ray luminosity
 of this source indicating that it radiates at a substantial fraction of the
 Eddington limit. Analysing the same {\sl ASCA} data, Reeves \& Turner
 (\cite{RT2000}) found also a significant emission line of ionized iron, but
 at a slightly smaller energy E=6.76$\pm$0.08\,keV
 (EW=274$^{+155}_{-140}$\,eV). In the present {\sl XMM-Newton} data analysis
 we found the presence of a very highly ionized line at E=6.7--7.0\,keV, 
corresponding to He-like iron (\ion{Fe}{xxv}) to H-like iron (\ion{Fe}{xxvi}). 
A broad or relativistic profile is statisticaly required and a very large EW
 is inferred (EW $\sim$ 500--900\,eV) .

\subsection{PG\,1202+282 (z=0.177)}

 The PG\,1202+282 data, in the 0.1--10\,keV energy band, were well represented 
 by an absorbed power law model ($\Gamma$=1.97$\pm$0.08) using {\sl ROSAT}
 plus {\sl BeppoSAX} data (Mineo et al. \cite{M2000}). In the present 
{\sl XMM-Newton} observation we find the clear presence of a soft excess 
below about 1.5\,keV. The broad-band continuum (0.3--12\,keV) is very well
 fitted either by a broken power law model, or by  a double Comptonization
 model.

\subsection{PG\,1244+026 (z=0.048)}

PG\,1244+026 is one of the five narrow line objects present in our sample, and
has the narrowest H$_{\beta}$ line (FWHM=830\,km\,s$^{-1}$). It has the lowest
 redshift and an extreme accretion rate relative to the Eddington rate.
PG\,1244+026 was previously observed with {\sl ROSAT} and {\sl ASCA} which 
revealed  rather steep 0.1--2\,keV and 2--10\,keV spectra with  photon indices
of 3.3$\pm$0.1 and 2.35$\pm$0.12, respectively (Fiore et al. \cite{F94},
 Fiore et al. \cite{F98}). A consistent value of the soft photon index was
 reported by Wang et al. (\cite{WBB96}) using {\sl ROSAT}. The present
 {\sl XMM-Newton} observation shows that the data are not well represented by 
a broken power law. In the soft X-ray range the large soft excess appears more
 curved than for the other objects. We find that the soft photon index is
 higher (steeper slope) than that observed with {\sl ROSAT}. The data are well
 fitted by  either a black body plus power law model, or by a double 
Comptonization model. The temperature inferred for the black body continuum
 (kT$\sim$197\,eV) is slightly larger than that expected for a standard
 accretion disk (kT$\sim$73\,eV). 

\subsection{PG\,1307+085 (z=0.155)}

From {\sl ROSAT} data, Wang et al. (\cite{WBB96}) found a soft photon index of
 3.37$\pm$0.13. It was also observed above 2\,keV by Lawson \& Turner 
(\cite{LT97}) using  {\sl Ginga}. They found that the spectrum was well
 represented by a simple power law over the 2--18\,keV energy range with a 
photon index of 1.90$\pm$0.11 with no detection of a narrow line at 6.4\,keV. 
 We find for the pn spectrum below 2\,keV a soft photon index of 
 2.22$\pm$0.05, smaller (flatter slope) than  {\sl ROSAT}. The pn spectrum
 (over the 2--12\,keV band) is well fitted by a power law with a photon index
 of 1.56$\pm$0.09 ($\chi^{2}$/dof=114.2/118, P($\chi^{2}$)=58.1$\%$). 
 This photon index appears harder than that found previously, however this 
could be due to  real hardening in the 2--12\,keV band or to the presence of a
 reflection component  below 12\,keV. We do not find any indication of a 
Fe\,K${\alpha}$ line, which appears to rule out the latter scenario. 

\subsection{PG\,1309+355 (z=0.184)}

PG\,1309+355 is one of the two radio-loud objects of our sample. 
 It was observed by {\sl ASCA} in 1997, according to the {\it Tartarus}
 database\footnote{http://tartarus.gsfc.nasa.gov/}, the spectrum was well
 fitted with a simple power law with a photon index of 1.77. This value is
 compatible with the one we find in the 2--5\,keV energy range with
 {\sl XMM-Newton}. We further detect an iron K$\alpha$ line near 6.4\,keV. 
 A broad line is only marginally required with a width of 
0.21$^{+0.24}_{-0.10}$\,keV.  If the line is broad, the inferred EW is much
 higher than the mean value observed for RLQ from {\sl ASCA} data,
 i.e. 85$\pm$15\,eV (Reeves \& Turner \cite{RT2000}).

\subsection{PG\,1322+659 (z=0.168)}

PG\,1322+659 was observed by {\sl ASCA} (George et al. \cite{G2000}) in May
 1999. The broad-band {\sl XMM-Newton} continuum is very well fitted by either
 a broken power law model, or the combination of a black body and a power law
 model, or by a double Comptonization model. No significant detection of an 
iron K${\alpha}$ line is required neither at 6.4\,keV nor at higher energy. 

\subsection{PG\,1352+183 (z=0.152)}

PG\,1352+183 was observed previously with {\sl BeppoSAX} by Mineo et al.
 (\cite{M2000}) who found that the data were well fitted by a single power law
 model. They also claimed the presence of a Fe\,K line at 6.43 $\pm$0.16\,keV,
with EW=760$\pm$460\,\,eV, but their detection was at only 1.5\,$\sigma$. With
 {\sl XMM-Newton}, we find the presence of a significant soft excess below
 2\,keV  with a soft X-ray photon index $\Gamma$=2.69$^{+0.10}_{-0.08}$. 
We find no Fe\,K line, however near 6\,keV the S/N is not high enough to
 detect a weak line. 

\subsection{PG\,1402+261 (z=0.164)}

PG\,1402+261 is a narrow-line quasar observed previously with {\sl BeppoSAX}
 by Mineo et al (\cite{M2000}). They found that the data were satisfactorily
 represented by a broken power law continuum with $\Gamma$=2.59$\pm$0.10 and 
1.52$\pm$0.30 for the soft and hard X-ray band, respectively. They did not
 report any detection of a Fe\,K line, however the MECS spectrum shows a large
 positive deviation above 7\,keV in the observer frame, compared to a power
 law continuum fit. PG\,1402+261 is one of the most interesting objects of the
 present  {\sl XMM-Newton} sample. It shows a very large positive deviation 
 from 5\,keV to 9\,keV (in observer frame). This could be interpreted as a
 very large Fe\,K line highly blueshifted, however other explanations, such as
 partial covering,  are not ruled out (Reeves et al. 2004, in preparation).

\subsection{Mrk 1383  (PG\,1426+015, z=0.086)}

Mrk 1383 was previously observed above 2\,keV with {\sl Ginga} in January 1991 
(Lawson \& Turner \cite{LT97}). The spectral fitting gave a photon index of
 2.07$\pm$0.20. We find here a 2--5\,keV photon index of 1.95$\pm$0.13 which
 is consistent with that found by {\sl Ginga}.

\subsection{PG\,1427+480 (z=0.221)}

PG\,1427+480 was observed by {\sl ROSAT} and the spectrum was well represented
 with a power law with  $\Gamma$=2.37$\pm$0.27. No observation above 2\,keV
 has been reported until now. Therefore the present {\sl XMM-Newton}
 observation allows to study the hard X-ray emission for the first time.  
 The source shows a soft X-ray excess and the soft X-ray photon index
 ($\Gamma$=2.52$^{+0.08}_{-0.03}$) is higher than the hard photon index
 ($\Gamma$=1.90$\pm$0.12). We  find neither the presence of a warm absorber 
 nor the presence of a significant Fe\,K line.

\subsection{Mrk 478 (z=0.077)}
 
Mrk 478 (PG\,1440+356) is one of the five narrow line objects of our sample. 
From {\sl ROSAT} observations, Wang et al. (\cite{WBB96}) inferred a soft
 photon index of 3.43$\pm$0.23. It has been observed with {\sl ASCA}
 (Vaughan et al. \cite{V99}) and a good fit was found using a combined black
 body/power law model with kT$_{\rm bb}$=89$\pm$0.12 and 
$\Gamma$=1.96$\pm$0.05. Reeves \& Turner (\cite{RT2000}) detected a soft
 excess and found a Fe\,K${\alpha}$ line at 6.37$\pm$0.07\,keV with
 EW=191$\pm$119\,eV (detection at 97$\%$). Mrk 478 was observed recently
 (August 8--9, 2000) using the  {\sl Chandra} Low Energy Transmission Grating
 spectrometer, and no significant emission or absorption feature were detected
 (Marshall et al. \cite{Ma2003}). Combining with {\sl BeppoSAX} data, 
they found that the spectrum is well fitted with two power laws with 
$\Gamma_{\rm soft}$=3.03$\pm$0.03 and $\Gamma_{\rm hard}$=1.4$\pm$0.2. They 
suggested that the X-ray continuum may result from Comptonization of disk 
thermal emission in a hot corona through a range of optical depths.  In the 
present {\sl XMM-Newton} data, we also find that  the spectrum is well fitted
 by the combination of a black body with a power law model, but it is much
 better fitted either by a broken power law model, or a double Comptonization
 model. Here we found $\Gamma_{\rm soft}$=3.12$^{+0.07}_{-0.06}$ 
 and $\Gamma_{\rm hard}$=2.45$\pm$0.13. The present soft spectrum has a
 similar shape as the one found with {\sl ROSAT} ($\Gamma$=3.43$\pm$0.23),
 while the hard spectrum appears to be slightly softer compared to the
 {\sl BeppoSAX/MECS} observation ($\Gamma$=2.19$\pm$0.05, 
Marshall et al. \cite{Ma2003}).

\subsection{PG\,1512+370 (z=0.371)}

PG\,1512+370  is one of the two radio-loud objects of our sample and has the
 highest redshift. The {\sl XMM-Newton} broad-band continuum is very well
 fitted either by a broken power law model, or by  a double Comptonization
 model. An iron  K$\alpha$ line is detected with a F-test probability of 
94.4$\%$. The energy of the line is 6.51$\pm$0.13\,keV which is compatible
 with neutral to moderately ionized iron. 

\subsection{Mrk\,876  (PG\,1613+658, z=0.129)}

Mrk\,876  was previoulsy observed above 2\,keV with {\sl Ginga} by 
Lawson \& Turner (\cite{LT97}). They found a photon index in the 2--18\,keV of
 1.63$\pm$0.03 which is rather flat for a RQQ quasar. No Fe\,K line was
 reported. In the present {\sl XMM-Newton} observation, we are able to
 constrain both the soft and hard X-ray slopes in this object. Mrk\,876 shows
 the presence of a soft excess and a warm absorber below 1\,keV, and no 
evidence for a Fe\,K line.

\subsection{PG\,1626+554 (z=0.132)}

PG\,1626+554 was observed by {\sl BeppoSAX} in February 1998 by Mineo et al.
 (\cite{M2000}). They found an unusual large excess below 0.2\,keV. They 
inferred that such an ``ultra-soft'' excess could be the  tail of a very steep
 unusually strong thermal component. Moreover, this ``ultra-soft'' excess
 could be related to the unusually flat optical to soft X-ray slope 
$\alpha_{\rm ox}$=$-$1.139 (Laor et al. 1997). In the present {\sl XMM-Newton}
 observation, we found that this object does not show a soft excess between
 0.3--2\,keV, but due to the present energy band we are not able to confirm 
 the ``ultra-soft'' excess found previously below 0.2 keV.

\end{document}